\documentclass[aps,prb,reprint,superscriptaddress,longbibliography]{revtex4-2}
\usepackage[english]{babel}
\usepackage[utf8]{inputenc}
\usepackage{amsthm}
\usepackage{mathtools}
\usepackage{physics}
\usepackage{xcolor}
\usepackage{graphicx,tikz}
\usepackage{adjustbox}
\usepackage{placeins}
\usepackage[T1]{fontenc}
\usepackage{csquotes}
\usepackage{comment}

\usepackage{tkz-euclide}

\usepackage{tikz}
\usetikzlibrary{patterns,arrows,calc,decorations.pathmorphing}

\usepackage{xr}

\usepackage[colorlinks=true, linkcolor=blue]{hyperref}
\hypersetup{citecolor=red}

\begin{document}
\title{Ultrasonic attenuation in inhomogeneous superconductors}

 \author{Aman Sardwal}
 \affiliation{Department of Physics, Indian Institute of Technology Madras, Chennai 600036, India}
 \affiliation{Center for Atomistic Modelling and Materials Design, IIT Madras, Chennai 600036, India}
 \author{Andreas Kreisel}
\affiliation{Department of Physics and Astronomy, Uppsala University, Box 524, 751 20 Uppsala, Sweden}
\affiliation{Niels Bohr Institute, University of Copenhagen, DK-2100 Copenhagen, Denmark}
\affiliation{Department of Physics, Indian Institute of Technology Madras, Chennai 600036, India}
 \affiliation{Quantum Centers in Diamond and Emergent Materials (QCenDiem)-Group, IIT Madras, Chennai 600036, India}
 \author{Shantanu Mukherjee}
 \affiliation{Department of Physics, Indian Institute of Technology Madras, Chennai 600036, India}
 \affiliation{Quantum Centers in Diamond and Emergent Materials (QCenDiem)-Group, IIT Madras, Chennai 600036, India}
 \affiliation{Center for Atomistic Modelling and Materials Design, IIT Madras, Chennai 600036, India}
% \date{\today} % Leave empty to omit a date
\begin{abstract}
Subgap bound states can produce pronounced features in the density of states without necessarily giving rise to a corresponding signal in ultrasonic attenuation. We show that the coherence factors act as a filter for the contribution of bound states to ultrasonic attenuation. In a phase-biased superconductor--normal--superconductor junction, the Andreev bound states produce a resonance in the attenuation when the phonon frequency matches the phase-dependent separation between the subgap levels. The situation is different at a $(110)$ grain boundary in a $d_{x^2-y^2}$ superconductor. There, nonmagnetic impurity-induced bound states remain weak in the attenuation despite their large spectral weight, because the relevant coherence factors nearly cancel. When local magnetism develops near the boundary, this cancellation is removed, the bound states become spin split, and a clear attenuation peak appears, which we argue is governed by a process involving emission of phonons by pair of quasiparticles at low temperatures. We obtain these results using a real-space Bogoliubov--de Gennes formulation of ultrasonic attenuation and benchmark the method against the known low-temperature behavior of homogeneous $s$- and $d$-wave superconductors. The results show that ultrasound can distinguish bound states with similar spectral signatures but different symmetry and spin structure.
\end{abstract}
\maketitle 

\section{Introduction}
\begin{figure}[t]
 \includegraphics[width=\linewidth]{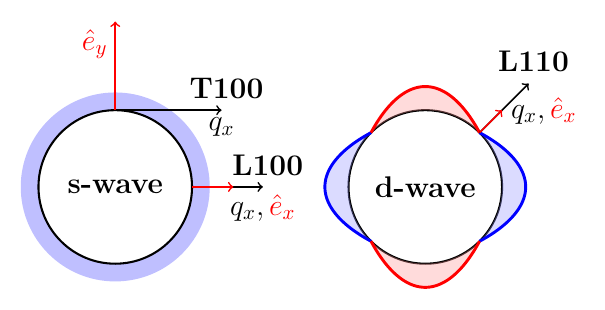}
\caption{\textbf{Example gap structures and phonon modes}. The blue (red) shaded regions denote areas where the superconducting gap is positive (negative). The relevant acoustic phonon modes—longitudinal and transverse—are shown with propagation along $\mathbf{q}$ and polarization along $\hat{\mathbf{e}}_\lambda$. Notation : Thkl, Lhkl denote transverse (T) and longitudinal (L) phonon modes respectively, propagating along hkl.}
  \label{fig:Schematic_sound_modes}
\end{figure}

Ultrasonic attenuation, the damping of sound waves as they propagate through a material, provides invaluable insights into material properties, particularly to detect and characterize phase transitions to superconductivity. This technique was among the first to confirm the conventional BCS theory~\cite{bommel,bardeen1957theory} for superconductivity, and since then, significant work has been conducted on the viscosity and elastic properties of such systems~\cite{Levy,Tsuneto,E_M_Forgan_1970,kadanoff1966ultrasonic,testardi,millis1988superconductivity,bruun1998acoustic}.
In superconductors, ultrasonic attenuation has been employed to investigate multiple superconducting phase transitions in $\text{UPt}_3$, and study gap structure in other unconventional superconductors~\cite{lin1992ultrasonic,batlogg1985lambda,bishop,muller1986observation}.The temperature dependence of the ultrasonic attenuation coefficient provides valuable information about low-energy quasiparticle excitations. In particular, its low-temperature power-law behavior can be used to infer the nodal structure of the superconducting order parameter~\cite{walker,morenocoleman,Rodriguez,Lupiennodes}. The location and symmetry of the gap nodes can be probed further by varying the polarization and propagation direction of the acoustic mode, as illustrated in Fig.~\ref{fig:Schematic_sound_modes}.
\\
For a given phonon mode, the electron--phonon matrix element may either remain finite or vanish at a superconducting gap node. Nodes at which the matrix element is finite are referred to as {active} nodes, whereas those at which it vanishes are termed {inactive} nodes. Consequently, different acoustic modes may exhibit distinct low-temperature attenuation behavior even for the same superconducting gap symmetry. For example, the L110 and L100 modes in a $d_{x^2-y^2}$-wave superconductor may couple differently to quasiparticles near the gap nodes, making a given node active for one mode and inactive for the other. At low temperatures, the attenuation associated with active nodes is enhanced by a factor proportional to $T^{-2}$ relative to that arising from inactive nodes~\cite{morenocoleman,walker}.

Recently, several intriguing experimental results have been reported for $\text{Sr}_2\text{RuO}_4$~\cite{Ghosh2021, Benhabib2021,sayakprb}, showing anomalous behavior in viscosity and elastic properties for specific strain modes. Notably, these experiments report an unexpected increase in viscosity for compressional strains below the superconducting transition temperature.
%Within the Landau-Ginzburg framework of phase transitions, this can be understood as a coupling between the multicomponent order parameter and the irreducible representations (irreps) of strain  that share the same symmetry channel. This coupling imposes constraints on the possible symmetry structure of the order parameter and gives us the most probable states possible in $\text{Sr}_2\text{RuO}_4$ ~\cite{Kivelson2020}
The proposed explanation for these anomalous results involves symmetry-preserved or accidentally degenerate chiral domains~\cite{Kivelson2020,PhysRevB.104.024511,joynt1986acoustic}, which were also previously predicted in Josephson interferometry experiments~\cite{yasui2020spontaneous}. Our work builds upon the same idea to explore the attenuation characteristics across different superconducting domains. These domains are anticipated to host Andreev bound states (ABS), which enhance the scattering cross-section for phonons in certain specific configurations, thereby leading to an increase in attenuation. This increase in attenuation may be relevant to the observed rise in viscosity below $T_c$ in ultrasound spectroscopy experiments on $\text{Sr}_2\text{RuO}_4$~\cite{sayakprb}. Furthermore, this study provides an opportunity for experimentalists to investigate the nodal structure of an unconventional superconductor, akin to the analysis performed in Josephson junction experiments.

\begin{figure*}[t]
 \includegraphics[width=\linewidth]{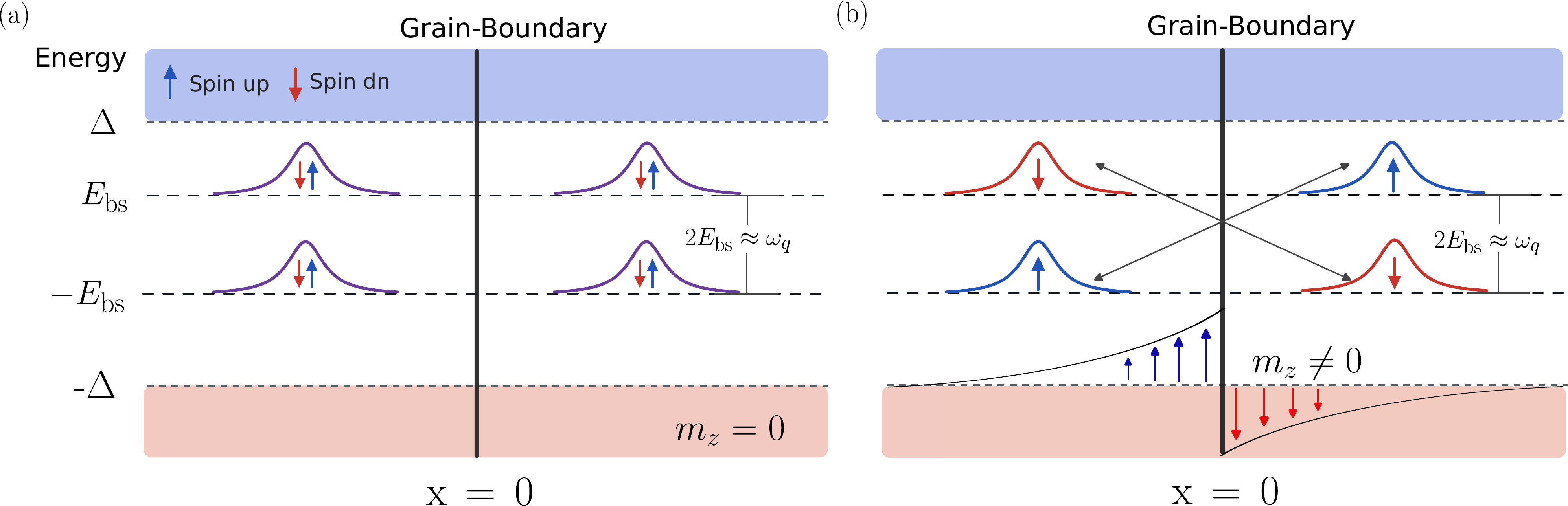}
\caption{\textbf{Attenuation from bound states
%Schematic illustration
}: {(a) Non-magnetic bound state: Andreev bound states and their quasiparticle spin content on either side of a grain boundary in an unconventional superconductor. (b) Magnetic bound state: In the presence of local magnetism, the spin-selective character of the Zeeman-split in-gap Andreev bound states changes sign across the grain boundary. The black arrows indicate the nonzero contribution to the ultrasonic attenuation coefficient from quasiparticles associated with the particle–hole-symmetric bound states.
}
}
\label{fig:Schematic_Spinsplitting}
\end{figure*} 
In this work, we consider the clean limit, in which the mean free path $\ell_e$ is larger than the phonon wavelength~\cite{morenocoleman}. We show that ultrasonic attenuation can serve as a useful probe for superconductors with broken translational symmetry. Ultrasonic attenuation measures the matrix element weighted transition susceptibility between Bogoliubov-de Gennes (BdG) quasi-particles. It can therefore distinguish systems with similar local density of states but different attenuation responses arising from variations in their local matrix element contributions.
In Sec.~\ref{sec:1}, we propose a method to determine the attenuation coefficient for a real-space inhomogeneous lattice using the equation of motion for the phonon number operator in the presence of electron-phonon coupling. 
This formalism is then benchmarked against well-established results for superconducting gap symmetries~\cite{walker}, as presented in Sec.~\ref{sec:2}. We subsequently generalize the framework to superconducting heterostructures, starting with SNS (Superconductor--Normal--Superconductor) junctions in Sec.~\ref{sec:3}. There, we investigate the dependence of the attenuation coefficient on the superconducting phase difference across the junction for different sound frequencies.
Finally, in Sec.~\ref{sec:4}, we study ultrasonic attenuation in $d$-wave superconductors with grain boundaries generated by impurity disorder~\cite{ohashi2001local,harter2007antiferromagnetic}. {We first consider non-magnetic impurity potentials that produce in-gap bound states at finite energies, and subsequently investigate the zero-energy bound-state regime, where Hubbard interactions induce local magnetism that splits the bound state.} While the bound-state splittings in these two cases can be of similar magnitude for appropriate impurity profiles and interaction strengths, we show that the attenuation coefficient provides a clear distinction between them. Specifically, local magnetism induces a spin-selective character of the in-gap bound states and leads to a detectable attenuation signature (see Fig.~\ref{fig:Schematic_Spinsplitting}).
\section{Methodology}
\label{sec:1}
As the sound wave is imposed on the sample, the electron-phonon interaction results in attenuation of the intensity of sound.
In the following, we set up a method to calculate the rate of attenuation described by the attenuation coefficient. We obtain it 
from the decay rate of the phonon number operator $\dot{N}_{\mathbf{q}\mathbf{\lambda}}=-{N}_{\mathbf{q}\mathbf{\lambda}} / {\tau}_{\mathbf{q}\mathbf{\lambda}}$, where $ {\tau}_{\mathbf{q}\mathbf{\lambda}} $ is phonon lifetime.
To model electron-phonon coupling, we start with a tight-binding Hamiltonian~\cite{walker},
\begin{equation}
\mathcal{H}=\sum_{\mathbf{ij} \sigma}t_{} (\mathbf{r}_{\mathbf{i}} - \mathbf{r}_{\mathbf{j}}) c^\dagger_\mathbf{i \sigma} c_\mathbf{j \sigma}
\label{eqn:ham_1}
\end{equation}
in which the hopping parameters $t_{} (\mathbf{r}_{\mathbf{i}} - \mathbf{r}_{\mathbf{j}})$ depend on the instantaneous position $\mathbf{r}_{\mathbf{i}}$ of ion placed at site \textbf{i}. The chemical potential is technically included as $\mu=-t_{} (\mathbf{r}_{\mathbf{i}} - \mathbf{r}_{\mathbf{i}})$.
Here, ${c^\dagger_\mathbf{i \sigma}, c_\mathbf{i \sigma}}$ are electron creation and annihilation operators at site $\mathbf{i}$ of spin $\sigma$. Expanding the hopping parameters about the equilibrium position of ions $\mathbf{r}_{0,\mathbf{i}}$  up to first order  reveals the electron-phonon coupling in its first order correction~\cite{walker}. The full second quantized Hamiltonian with electron-phonon interaction then consists of following terms,
\begin{align}
\mathcal{H}_{ph}&=\sum_{\mathbf{q} \lambda}\hbar \omega_{\mathbf{q},\lambda}\left(b_{\mathbf{q} \lambda}^\dagger b_{\mathbf{q} \lambda} +\frac{1}{2}\right)\\
\mathcal{H}_{el}&=\sum_{\mathbf{ij} \sigma} t({\mathbf{r}}_{0,\mathbf{i}} - {\mathbf{r}}_{0,\mathbf{j}}) c^\dagger_{\mathbf{i}\sigma} c_{\mathbf{j}\sigma}\\
\mathcal{H}_{el-ph}&=\sum_{\mathbf{ij} \sigma} ({\nabla} t_{\mathbf{ij}} \cdot (\mathbf{u}_{\mathbf{i}} - \mathbf{u}_{\mathbf{j}})) c^\dagger_{\mathbf{i}\sigma} c_{\mathbf{j}\sigma}+ \dots
\label{eq:Full_Hamiltonian}
\end{align}
Here, the first term represents the phonon contribution to the Hamiltonian, where
$b_{\mathbf{q}\lambda}^{\dagger}$ and $b_{\mathbf{q}\lambda}$ denote the phonon creation and
annihilation operators, respectively, for a phonon mode with wavevector $\mathbf{q}$, 
polarization $\hat{\mathbf{e}}_{\lambda}$ and frequency $\omega_{\mathbf{q},\lambda}$. The second term corresponds to the standard
tight-binding electronic Hamiltonian. The third
term describes the electron--phonon interaction, where $\mathbf{u}_\mathbf{i}$ denotes the displacement of the ion $\mathbf{i}$ from its equilibrium position, given by
$\mathbf{u}_{\mathbf{i}}={\mathbf{r}}_{\mathbf{i}}-{\mathbf{r}}_{0,\mathbf{i}}$ and ${\nabla} t$ represents the gradient of the hopping parameter with respect to the displacement.
%$ {\mathcal{H}}^{\prime} $.
To incorporate a phonon mode of momentum ${\textbf{q}}$ and polarization $\hat{\mathbf{e}}_{\lambda}$, we perform Fourier transformation over the ionic displacement $\mathbf{u}_\mathbf{i}$ in terms of phonon creation and annihilation operator as, 
\begin{align}
\mathbf{u}_\mathbf{i} =  \sum_{\mathbf{q}\lambda}\frac{\ell_{\mathbf{q}\lambda}}{\sqrt{2N}}\left(b_{-\mathbf{q}\lambda}^{\dagger}+b_{\mathbf{q}\lambda}\right)\hat{\mathbf{e}}_{\lambda}e^{i\mathbf{q}\cdot\mathbf{r_i}}
\end{align}
where $\ell_{\mathbf{q}\lambda} = \sqrt{\frac{\hbar}{M \omega_{\mathbf{q}\lambda}}}$ is zero-point displacement amplitude, $N$ is the number of ions, and $M$ is the ionic mass. The corresponding displacement difference between sites $\mathbf{i}$ and $\mathbf{j}$ is then
\begin{align}
\mathbf{u}_\mathbf{i}-\mathbf{u}_\mathbf{j} =  \frac{1}{\sqrt{2N}}\sum_{\mathbf{q} \lambda }& \ell_{\mathbf{q}\mathbf{\lambda}}\left\{  b_{-\mathbf{q}\mathbf{\lambda}}^{\dagger}\left(e^{i\mathbf{q}\cdot\mathbf{r_i}}-e^{i\mathbf{q}\cdot\mathbf{r_j}} \right)\right.\notag\\
&\left.+b_{\mathbf{q}\mathbf{\lambda}}\left(e^{i\mathbf{q}\cdot\mathbf{r_i}}-e^{i\mathbf{q}\cdot\mathbf{r_j}} \right)\right \}\hat{\mathbf{e}}_{\lambda}.
\end{align}
Thus, the electron--phonon interaction Hamiltonian can be expressed in terms of the phonon normal modes as 
\begin{align}
\mathcal{H}_{el-ph}&=\sum_{\substack{\mathbf{ij}  \sigma\\ \mathbf{q} \lambda}} \frac{\delta {t}_{\mathbf{i} \mathbf{j}}^\lambda \ell_{\mathbf{q} \lambda} \phi_{\mathbf{i} \mathbf{j}}\left(\mathbf{q}\right) }{\sqrt{2N}}\left(b_{-\mathbf{q} \lambda}^\dagger + b_{\mathbf{q}\lambda} \right)c^\dagger_{\mathbf{i}\sigma} c_{\mathbf{j}\sigma}
\label{eq:Hep}
\end{align}
where $\delta t_{\mathbf{ij}}^\lambda={\nabla} t_{\mathbf{ij}} \cdot\hat{\mathbf{e}}_{\lambda}$ is the electron-phonon coupling matrix (see Appendix~\ref{app:2} for more details) and $\phi_{\mathbf{i} \mathbf{j}}\left(\mathbf{q}\right)=e^{i\mathbf{q}.\mathbf{r}_\mathbf{i}}-e^{i\mathbf{q}.\mathbf{r}_\mathbf{j}}$.

The time derivative $\dot{N}_{\mathbf{q}\mathbf{\lambda}}$ of the phonon density can be obtained using Heisenberg's equation of motion,
\begin{align}
&{i}\hbar \dot{N}_{\mathbf{q} \lambda}(t) = \left[b^\dagger_{\mathbf{q} \lambda}b_{\mathbf{q} \lambda}, \mathcal{H}_{0}+\mathcal{H}_{el-ph}\right] \notag \\
&= \sum_{\mathbf{ij}\sigma} \frac{\delta t_{\mathbf{ij}}^ \lambda \ell_{\mathbf{q} \lambda}}{\sqrt{2N}} \left(  b_{\mathbf{q} \lambda}^{\dagger} \phi_{\mathbf{i} \mathbf{j}}^*\left(\mathbf{q}\right) -   b_{\mathbf{q} \lambda} \phi_{\mathbf{i} \mathbf{j}}\left(\mathbf{q}\right) \right) c^\dagger_{\mathbf{i}\sigma} c_{\mathbf{j}\sigma}
\label{eqn:NqH0}
\end{align}   
For the calculation of the attenuation coefficient, we evaluate the first order correction in ${\dot{N}_{\mathbf{q}\mathbf{\lambda}}}$ using linear response theory,
\begin{equation}
\left\langle{\dot{N}_{\mathbf{q}\mathbf{\lambda}}}(t)\right \rangle =-\frac{{i}}{\hbar} \int_{0}^{\infty} \mathrm{d} t^\prime \left\langle\left[{{\dot{N}}^{\text{I}}_{\mathbf{q}\mathbf{\lambda}}}(t^\prime), {\mathcal{H}_{el-ph}}\right]\right\rangle_{\!0}.
\label{eqn:kuboNq}
\end{equation}
Here ${\dot{N}}^{\text{I}}_{\mathbf{q}\mathbf{\lambda}}$ is the interaction picture corresponding to Heisenberg operator $\dot{N}_{\mathbf{q}\mathbf{\lambda}}$, which gets time-evolved with respect to the unperturbed Hamiltonian $\mathcal{H}_0$ (involves both $ \mathcal{H}_{el}$ and $ \mathcal{H}_{ph}$). $\left\langle \hspace{2mm} \right\rangle_{\!0}$ indicates the expectation values, taken w.r.t the unperturbed Hamiltonian ground state.
The commutator in Eq.~\eqref{eqn:kuboNq} can be expressed as,
% \begin{equation}
\begin{align}
& \left\langle\left[{ {\dot{N}}^{\text{I}}_{\mathbf{q}\mathbf{\lambda}}}(t), {\mathcal{H}}_{el-ph}\right]\right\rangle_{\!0}=- i \frac{\ell_{\mathbf{q}\mathbf{\lambda}}^2N_{\mathbf{q}\mathbf{\lambda}}}{2 \hbar N}\sum_{\mathbf{ij} \sigma}\sum_{\mathbf{i^\prime j^\prime}\sigma'} {\delta t_{\mathbf{i}\mathbf{j}}^\lambda\delta t_{\mathbf{i}^\prime \mathbf{j}^\prime}^\lambda } \notag \\
&\! \left( \beta_{\mathbf{i j i^\prime j^\prime} } (\mathbf{q}) e^{i\omega_\mathbf{q} t}- \beta_{\mathbf{i j i^\prime j^\prime}}^* (\mathbf{q})e^{-i\omega_\mathbf{q} t}\right)\!
\langle\left[ c_{\mathbf{i}\sigma}^{\dagger}(t) c_{\mathbf{j}\sigma}(t),c^\dagger_{\mathbf{i}^\prime\sigma'} c_{\mathbf{j}^\prime \sigma'}\right]\rangle
\label{eqn:finalcom}
\end{align}
where we defined the phases $\beta_{\mathbf{ij} \mathbf{i^\prime j^\prime}}\left(\mathbf{q}\right) = \beta_{ \mathbf{i^\prime j^\prime} \mathbf{ij}} ^* \left(\mathbf{q}\right)= \phi_{\mathbf{i} \mathbf{j}}^* \left(\mathbf{q}\right) \phi_{\mathbf{i}^\prime \mathbf{j}^\prime}\left(\mathbf{q}\right)$.
Since the phonon operators evolve under \(\mathcal{H}_{{ph}}\) in their interaction-picture representation, the time dependence for phonon creation and annihilation operator is $e^{\pm{i} \omega_{\mathbf{q}} t } $, respectively. The final expression for the change in $\left\langle\dot{N}_{\mathbf{q}\mathbf{\lambda}}(t)\right \rangle$ can be obtained by substituting  Eq.~\eqref{eqn:finalcom} into Eq.~\eqref{eqn:kuboNq}. The phonon relaxation rate, $1/\tau_{\mathbf{q}\lambda}$, is identified from the first-order correction to $\langle \dot{N}_{\mathbf{q}\lambda}(t)\rangle$. The corresponding attenuation coefficient is given by $\alpha(\mathbf{q},\lambda)=\left[v_s\tau_{\mathbf{q}\lambda}\right]^{-1}$, where $v_s$ is the sound velocity, resulting in
\begin{align}
\alpha\left({\mathbf{q},\mathbf{\lambda}}\right)&=\frac{\ell_{\mathbf{q}\mathbf{\lambda}}^2}{{2\hbar ^2 v_s N}}\sum_{\substack{\mathbf{i},\mathbf{j},\sigma \\ \mathbf{i}',\mathbf{j}',\sigma'}} \int_{0}^{\infty} d t  \left\langle\left[ c_\mathbf{i \sigma}^{\dagger}(t) c_\mathbf{j \sigma}(t)\, , \, c^\dagger_{\mathbf{i}^\prime \sigma^\prime} c_{\mathbf{j}^\prime \sigma^\prime}\right]\right\rangle_{\!0}  \;\,     \notag \\ 
& \times \quad  \delta t_{\mathbf{ij}}^\lambda\delta t_{\mathbf{i}^\prime \mathbf{j}^\prime}^\lambda   \left( \beta_{\mathbf{i j i^\prime j^\prime} }(\mathbf{q}) e^{i\omega_q t}-\beta_{\mathbf{i j i^\prime j^\prime}}^* (\mathbf{q}) e^{-i\omega_q t}\right)
\label{eqn:alphaSCDraft}  
\end{align} 
This quantity represents the attenuation coefficient $\alpha\left({\mathbf{q},\mathbf{\lambda}}\right)$ for a given phonon mode labeled by momentum $\mathbf{q}$ and polarization $\lambda$. The important step is that the commutator in this equation
can be evaluated most conveniently in the eigenbasis of the bare Hamiltonian $\mathcal{H}_0$. Expressing all operators in this basis allows one to perform the time evolution inside the commutator explicitly.
As a concrete example, consider a translationally invariant electronic Hamiltonian. Its eigenstates are labeled by the crystal momentum $\mathbf{k}$ and spin $\sigma$, and the real-space fermionic operators can be expanded as $c_{\mathbf{i} \sigma} = \sum_{\mathbf{k}} e^{i\mathbf{k}\cdot\mathbf{r}_{\mathbf{i}}}\, c_{\mathbf{k}\sigma}$,
where $c_{\mathbf{k}\sigma}$ annihilates a fermion in the momentum eigenstate. In this basis the Hamiltonian takes a diagonal form, $ \mathcal{H}_{el} = \sum_{\mathbf{k}\sigma} \xi_{\mathbf{k}\sigma}\, c_{\mathbf{k}\sigma}^\dagger c_{\mathbf{k}\sigma}$ and the Heisenberg evolution of the operators is given by $c_{\mathbf{k}\sigma}(t)
= c_{\mathbf{k}\sigma}\, e^{-i\xi_{\mathbf{k}\sigma} t /\hbar}$. Carrying out this transformation for all operators appearing in the commutator of  Eq.~\eqref{eqn:alphaSCDraft} allows us to rewrite the r.h.s. in terms of the eigenstates and eigenvalues of $\mathcal{H}_0$. This facilitates an explicit evaluation of the attenuation coefficient $\alpha_{\mathrm{n}}\left({\mathbf{q},\mathbf{\lambda}}\right)$ in the normal state in terms of the electronic wavefunctions and the electron-phonon matrix elements.
\begin{figure*}[t]
\centering
\includegraphics[width=\linewidth]{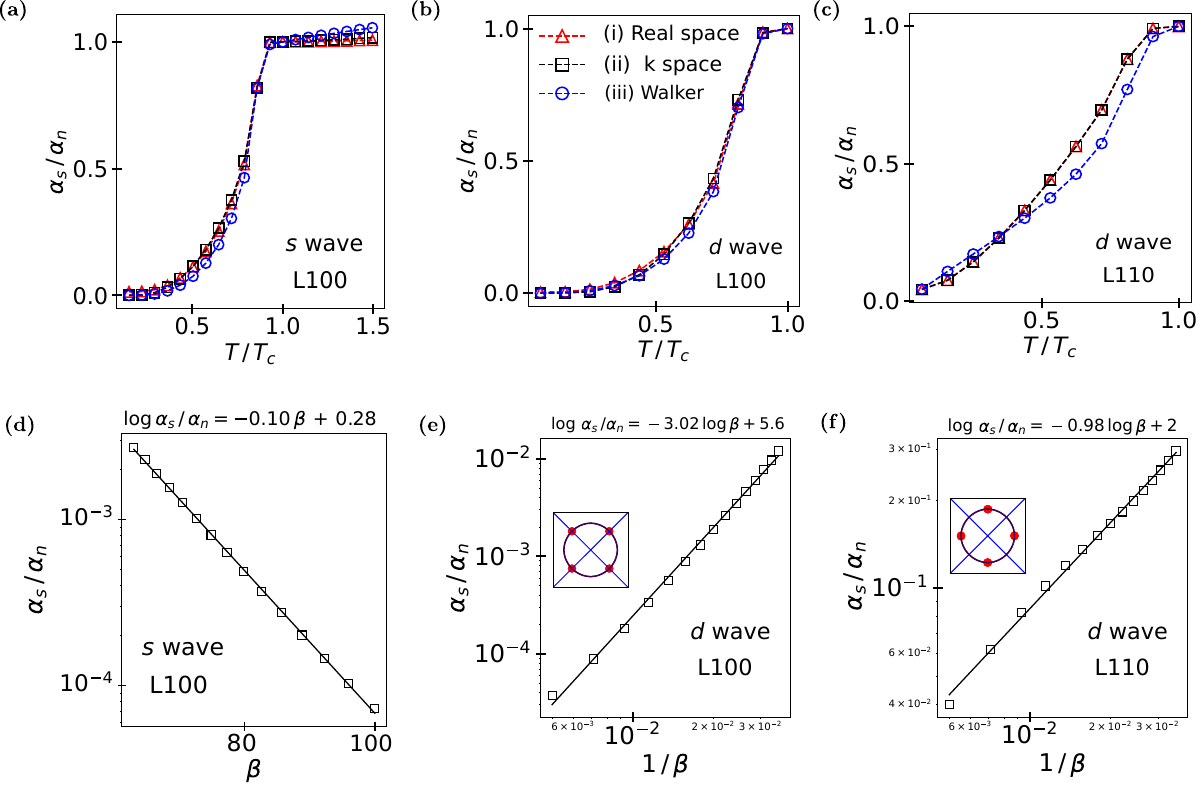}
\caption{Temperature dependence of the normalized attenuation coefficient $\alpha_s$ in the superconducting state relative to the normal state.  
(a) Attenuation for \text{$s$-wave} superconductor ($\Delta (\mathbf{k}) = \Delta_0$) in the L100 mode. %, comparing results from \text{real-space calculations, full k-space integration, and the analytical expression from} ~\cite{walker}.  
Attenuation for a \text{$d$-wave} superconductor ($\Delta (\mathbf{k}) = \Delta_0 (\cos k_x - \cos k_y)$) are shown for L100 mode (b), and L110 mode (c).  
Panels (d), (e)  and (f) correspond to the low temperature characteristics of the (a), (b) and (c), respectively showing the exponential characteristics for $s$-wave and power law for $d$-wave gap functions. Inset in panel (e),(f) show the Fermi surface (black solid line) with $n=0.8$ and $t'=0.3$. The nodal lines of superconducting gap (blue) and zeros of electron-phonon interaction (red dots) are also shown. } %Inset  (e) displays the Fermi surface (black curve), the zeros of the electron–phonon matrix element $F(\mathbf{k},\mathbf{q})$ (red dots; see Ref.~\cite{walker} and the Supplementary Table), and the nodal lines of the $d_{x^2 - y^2}$ gap function (Blue lines). 
\label{fig:benchmarking}
    \end{figure*}
\begin{figure}[t]
 \includegraphics[width=\linewidth]{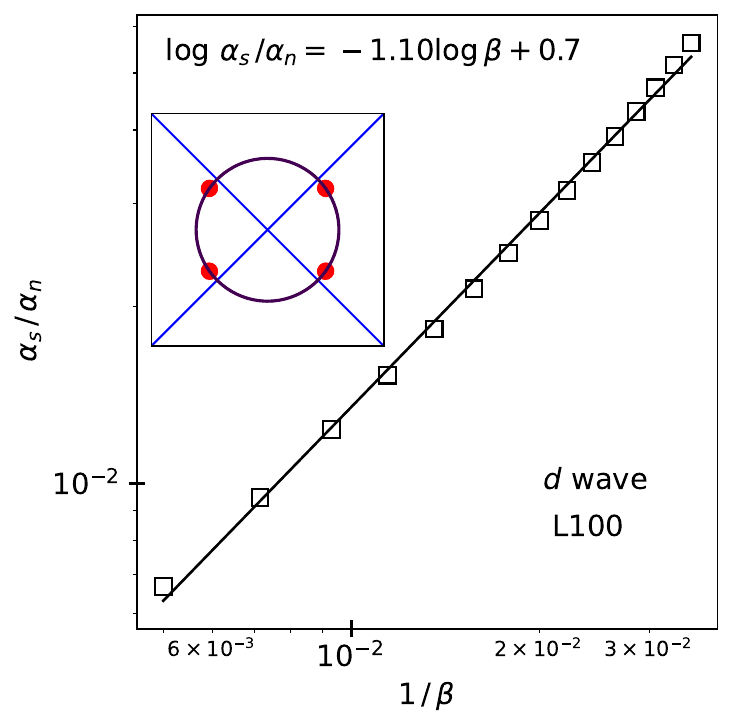}
\caption{The low temperature characteristics of   $d_{x^2 - y^2}$ for L100 mode for a different Fermi surface ($n=0.6$), with zeros of the electron–phonon matrix element $F(\mathbf{k},\mathbf{q})$, not lying along the nodal lines.}
  \label{fig:InactiveVsactiveNodes}
\end{figure} 

To study a superconducting system, the unperturbed Hamiltonian must describe the system in its superconducting state. Within mean-field theory, the Hamiltonian $\mathcal{H}_0$ then acquires the additional terms from the superconducting phase
\begin{align}
    \mathcal{H}_{sc}=\sum_{\mathbf{ij}}\left(\Delta_{\mathbf{ij}} c_{\mathbf{i} \uparrow}^{\dagger} c_{\mathbf{j} \downarrow}^{\dagger}+\Delta_{\mathbf{ij}}^* c_{\mathbf{j} \downarrow} c_{\mathbf{i} \uparrow}\right),
    \label{eq:base_sc_ham}
\end{align}
where for an even parity superconducting state the gap function between bonds connecting sites $i$ and $j$ can be calculated as
 \begin{align}  
 \Delta_{\mathbf{ij}}=\frac{V_{\mathbf{ij}}}{2}\left(\left\langle c_{\mathbf{i} \uparrow} c_{\mathbf{j} \downarrow}\right\rangle-\left\langle c_{\mathbf{i} \downarrow} c_{\mathbf{j} \uparrow}\right\rangle\right)
\end{align}
where  $V_{\mathbf{ij}}$ is the interaction parameter. This mean-field superconducting Hamiltonian can be diagonalized using a Bogoliubov-de Gennes (BdG) transformation. To evaluate Eq.~\eqref{eqn:alphaSCDraft} in real space, we use the BdG transformation,
\begin{align}  
{c}^\dagger_{\mathbf{i} \sigma}=\sum_{n >0 } {u_{\mathbf{i} \sigma}^{n *}}^{}{\gamma}^\dagger_{{n} }-{\sigma}v_{\mathbf{i} \sigma}^{n }{\gamma}_{n}
\label{eq:bdgtransformation}
\end{align}
Here ${\gamma}^{\dagger}_{n}$ denotes the fermionic creation operator for a Bogoliubov quasiparticle with energy $E_{n}$, and the summation runs over the positive eigenvalues of the Bogoliubov--de Gennes spectrum. The coefficients $u$ and $v$ are the eigencomponents of the superconducting Hamiltonian introduced above. For the annihilation operator $c_{\mathbf{i}\sigma}$, the corresponding transformation is simply the Hermitian conjugate of Eq.~\eqref{eq:bdgtransformation}. As shown in the Appendix~\ref{app:1}, one can then rewrite the expression for the attenuation rate in the superconducting state as,
    \begin{align}
\alpha_{\mathrm{s}}\left({\mathbf{q},\mathbf{\lambda}}\right)= - &  \sum_{m,n } \frac{\ell_{\mathbf{q}\lambda}^2 |\mathcal{F}_{mn}^\lambda(\mathbf{q}) |^2}{2 {\hbar v_s N}}\mathrm{Im}[\boldsymbol{\chi}_{\mathbf{q}}^{mn} (\omega_{\mathbf{q}\lambda})] 
\label{eq:alpha_sc_final}
    \end{align}
where the sum of the ($n,m$) indices now runs across the entire positive and negative BdG spectrum. Coherence factor weighted electron--phonon matrix elements $ \mathcal{F}_{mn}^\lambda(\mathbf{q})$ is given by
\begin{align}
 \mathcal{F}_{mn}^\lambda(\mathbf{q})&=\sum_{\mathbf{ij} \sigma}\delta t_{\mathbf{ij}}^\lambda \phi_{\mathbf{i} \mathbf{j}}^*(\mathbf{q})\left( {u_{\mathbf{i} \sigma}^{m *}} {u_{\mathbf{j} \sigma}^{n }}  -{v_{\mathbf{i} \sigma}^{n }} {v_{\mathbf{j}\sigma}^{m *}} \right)
\end{align}
and the corresponding $\boldsymbol{\chi}_{\mathbf{q}}^{mn} (\omega_{\mathbf{q}\lambda})$ is
\begin{align}
  \boldsymbol{\chi}_{\mathbf{q}}^{mn} (\omega_{\mathbf{q}\lambda})
&=\frac{
 f(E_m) - f(E_n) 
}{
\hbar \omega_{\mathbf{q}\lambda} + E_m - E_n + i\eta
},
 \label{eq:coherencefactors}
\end{align}
here, $E_m$ and $E_n$ are the BdG eigenvalues, 
$f(E)$ is the Fermi–Dirac distribution function, and $\eta$ is a small broadening parameter. In the following, we refer to
\begin{align}
\chi_{\mathbf{q}}=\sum_{m,n}\boldsymbol{\chi}_{\mathbf{q}}^{mn} (\omega_{\mathbf{q}\lambda})
\label{eq:Lindhard_function}
\end{align}
as the susceptibility function. However, physical susceptibility also includes the corresponding matrix-element contributions.
Our real space calculations are based on an effective mean-field model for a superconductor, using a ribbon configuration where the system is periodic along the $y$ direction. This allows us to model the homogeneous superconductor, and inhomogeneous systems like the SNS junction and the grain boundary configuration with reasonably large system sizes. The BdG Hamiltonian for such superconducting systems is
\begin{align}
\mathcal{H}_{0}&=\sum_{i j, \sigma , k_y} t_{i, j } (k_y)  c_{i \sigma}^{\dagger}(k_y) c_{j \sigma^{}} (k_y)  - \mu  c_{i \sigma}^{\dagger}(k_y) c_{i \sigma^{}} (k_y) \notag \\
& + \sum_{i j k_y}\left[\Delta_{i j} (k_y)c_{i \uparrow}^{\dagger}(k_y) c_{j \downarrow}^{\dagger}(-k_y) + h.c
\right].
\label{eq:benchmarkHamiltonian}
\end{align}
Here, $i$ and $j$ denote the $x$-coordinates on square lattice sites, and $k_y$ represents the Bloch momentum along the $y$-direction. The superconducting gap order parameter, $\Delta_{ij}$ is assumed to correspond to an even parity superconductor, that is either onsite ($s$-wave) or a nearest-neighbor bond-dependent $d_{x^2-y^2}$-wave symmetry, respectively. The chemical potential $\mu$, and $\Delta_{ij}$ are determined using the self-consistent gap equation.  
\begin{figure}[b]
\includegraphics[width=\linewidth]{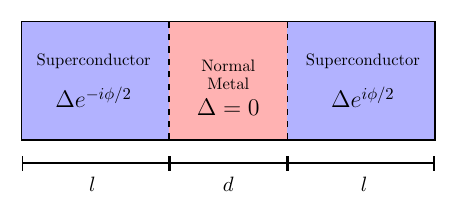} 
\caption{Schematic of a SNS junction with phase difference of $\phi$ between the left and right superconductor.}
\label{fig:SNS_junction}
\end{figure}
\begin{figure}[t]
    \includegraphics[width=\linewidth]{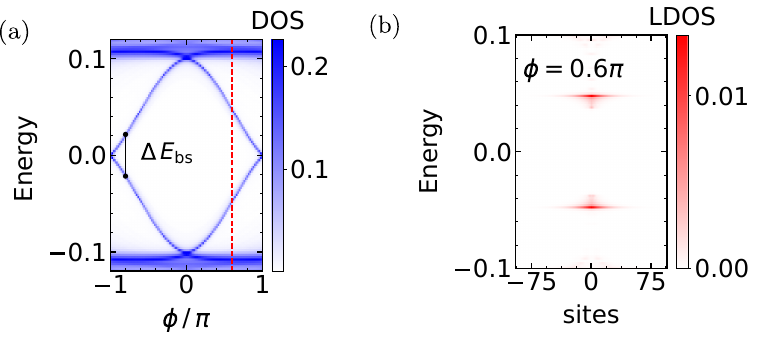}
    \caption{(a) Density of States of the SNS junction as a function of the phase difference $\phi$ across the junction showing the in-gap Andreev bound states. %The plot shows the formation of pairs of in-gap Andreev bound states whose energy difference $\Delta E_{bs}$ (denoted by black line) depends on $\phi$. 
    (b) Local Density of States across the junction for $\phi \approx 0.6 \pi$  (red dashed line in (a)), highlighting the spatial extension of subgap states.
        % (c) and (d) shows the variation of the attenuation coefficient $\alpha_{\mathbf{q}}$ with phase difference $\phi$ for different values $\omega_q$. 
    }
    \label{fig:DOS_ABS}
\end{figure}
% %%%%%%%%%%%%%%%%%%%%%%%%%
%%%%%%%%%%%%%%%%%%%%%%%%%%
\section{Benchmarking}
\label{sec:2}
In this section, we evaluate the ultrasonic attenuation coefficient for various superconducting gap symmetries and phonon polarizations in a homogeneous system. We reproduce the established contrast between fully gapped $s$-wave and nodal $d$-wave superconductors and benchmark it with three complementary formulations: (i) the real space calculation done in ribbon configuration based on Eq.~\eqref{eq:alpha_sc_final}, with real space along $x$ and momentum space along $y$; (ii) a fully momentum space formulation obtained by Fourier transforming the homogeneous real-space model; and (iii) the analytical treatment of Walker \textit{et al.}~\cite{walker}, which neglects processes involving the creation or annihilation of two BdG quasiparticles (see Appendix~\ref{app:1}). The nodal and anisotropic structure of the $d$-wave gap produces a pronounced distinction between the attenuation of the $(100)$ and $(110)$ phonon modes. This directional dependence is absent for the isotropic $s$-wave state. 
 
\paragraph*{Model parameters.}For the electronic structure of the normal state, we choose the next-nearest-neighbor (nnn) hopping as $t^\prime=0.3 t$ and fix
the total electron density at $n = 0.8$. In the following, we represent all energy scales in units of nearest-neighbor hopping by choosing $t=1$ that approximately corresponds to an energy of around $400$ meV in real materials~\cite{walker}. 
The superconducting pairing strengths are chosen as $V_{\mathbf{ij}} = 0.825 \delta_{\mathbf{ij}}$ for the 
$s$-wave case and $V_{\mathbf{ij}} = 0.78$ for the $d$-wave case (${\mathbf{i,j} \in \text{n.n}}$), which yield a superconducting order parameter of approximately $\Delta_{\mathbf{ij}} \approx 
0.1~\text{}$ for $s$-wave pairing and $\Delta_{\mathbf{ij}} \approx 
0.08~\text{}$ for $d$-wave pairing, respectively. For the attenuation calculation, the broadening factor was taken as $\eta \approx 0.005$. \\
A longitudinal phonon mode L100 contains both polarization and propagation directed along the $x$-axis. 
The electron-phonon coupling is assumed to be proportional to the underlying hopping parameters, corresponding to an exponential decay of the hopping parameters over a characteristic length scale $\zeta$. For the homogeneous superconductor the coupling coefficient $\delta t_{\mathbf{ij}}^{\lambda}$ is assumed to be nonzero for both nearest-neighbor (nn) bonds, with $\mathbf i = (x,y)$ and $\mathbf j = (x\pm 1, y)$, as well as for next-nearest-neighbor (nnn) bonds, with $\mathbf i = (x,y)$ and $\mathbf j = (x \pm 1, y \pm 1)$. For the junction and grain boundary problems discussed below, longer range tight binding hopping matrix elements are considered in order to allow electron-phonon interaction matrix elements to couple opposite sides of the junction region. These longer range hoppings are substantially reduced and have a negligible effect on the electronic structure, but as discussed below, play a critical role in the attenuation behavior near the SNS junction and across grain boundaries (see Appendix~\eqref{app:2} for more details).
\\
Figures~\ref{fig:benchmarking}(a)--(c) compare our formalism (i) with methods (ii) and (iii) for the $s$-wave L100 mode, the $d$-wave L100 mode, and the $d$-wave L110 mode. We then performed a low-temperature analysis of these cases. For an $s$-wave superconductor, the attenuation coefficient exhibits an exponential suppression at low temperatures, see Fig.~\ref{fig:benchmarking}d. In contrast, for a $d$-wave order parameter, the low-temperature behavior depends on whether the phonon mode couples to active or inactive nodes. For the Fermi surface considered in Fig.~\ref{fig:benchmarking}, the coupling $\delta t_{\mathbf{ij}}$ associated with the L100 mode vanishes at the nodal points, making these nodes inactive. This can be seen from the inset of Fig.~\ref{fig:benchmarking}(e), which shows the Fermi surface (black), the nodal lines of the superconducting gap (blue) and the zeros of the electron-phonon coupling matrix element (red). This fine-tuned cancelation can be lifted by modifying the Fermi surface, which shifts the zeros of $\delta t_{\mathbf{ij}}$ and reactivates the nodal contribution. In order to highlight the comparison with an active node scenario, in Fig.~\ref{fig:InactiveVsactiveNodes} we show the low temperature characteristics for a different doping (black curve in the inset of Fig.~\ref{fig:InactiveVsactiveNodes}, having $n=0.6$) such that the nodes (blue lines in the inset) do not align with zeros of the elements of the electron-phonon coupling matrix (red dots in the inset), making the nodes active.
As a consequence, the attenuation coefficient associated with the active nodes of L100, shown in Fig.~\ref{fig:InactiveVsactiveNodes}, exhibits a characteristic dependence of $T^{1.1}$ at low temperatures. This behavior is in sharp contrast to that of the inactive nodes, whose low-temperature attenuation is proportional to $T^{3.02}$, as shown in Fig.~\ref{fig:benchmarking}e. The distinct temperature exponents reflect the fundamentally different low-energy quasiparticle excitations probed by active and inactive nodal directions, which is the extra factor of $1/T^2$ for active nodes~\cite{morenocoleman}.\\
An L110 longitudinal phonon mode with polarization directed along the $(110)$ direction is also considered. To correctly implement periodic boundary conditions, the system must be transformed to a rotated basis, $(x,y) \rightarrow (x+y,\, -x+y)$  (see Fig.~\ref{fig:lattice_lineimpurity} for a schematic), under which the lattice periodicity changes from $a$ to $\sqrt{2}\,a$. This transformation ensures periodicity along the $x = -y$ direction (-110) , which is perpendicular to the phonon propagation direction (110). For  an $s$-wave superconductor, the attenuation coefficient retains its exponentially suppressed low-temperature behavior, reflecting the fully isotropic gap. In contrast, for a $d$-wave order parameter, the coupling $\delta t_{\mathbf{ij}}$ remains finite at the nodal points (Fig.~\ref{fig:benchmarking}(f) inset), confirming that the nodes are active~\cite{walker,morenocoleman}. In this case, the ultrasonic attenuation shows an approximately linear temperature dependence, as shown in Fig.~\ref{fig:benchmarking}(f). Its inset shows that the nodal lines (blue) are anti aligned with respect to the zeros of electron-phonon coupling at the gap maxima (red), making the attenuation rise more prominently as they are not being suppressed by the electron-phonon coupling matrix elements. In general, we find good agreement between the various approaches for calculating the ultrasonic attenuation. 
\begin{figure}[t]
    \includegraphics[width=\linewidth]{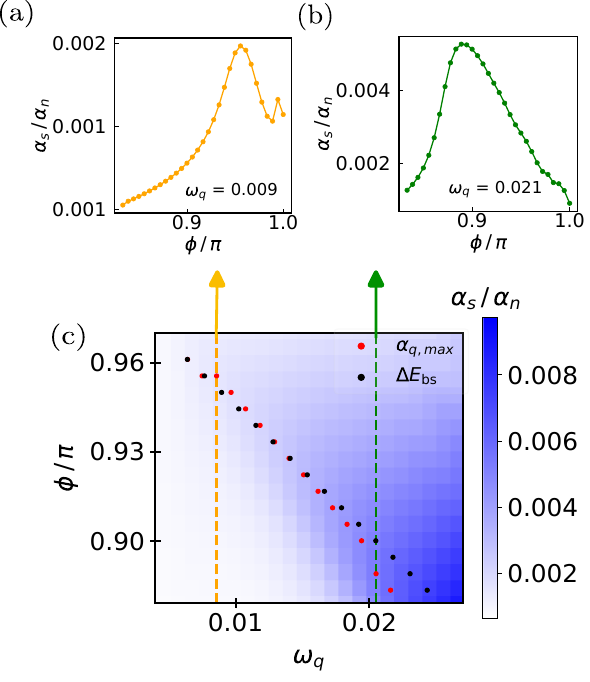}
    \caption{Variation of $\alpha_{\mathrm{s}}/\alpha_{\mathrm{n}}$ with sound frequency $\omega_q$ for different values of the phase difference $\phi$. The ABS appear in pairs, with their energy difference $\Delta E_{bs}$ (see Fig.~\ref{fig:DOS_ABS}(a)) exhibiting a well-defined relationship with the phase difference $\phi$, represented by black dots in panel (c). This is compared with the values of $\phi$ at which the attenuation peak occurs for a given $\omega_q$ (red dots). Panels (a) and (b) show cuts through the data for two specific values of $\omega_q$ (0.009 and 0.021), represented by yellow and green vertical lines respectively.}
    \label{fig:SNS}
\end{figure}
\section{Attenuation in SNS Junctions }
\label{sec:3}

In this section, we present our analysis of ultrasonic attenuation in a 
superconductor-normal-superconductor (SNS) junction for the L100 phonon mode. 
The SNS junction consists of left and right superconducting 
regions, maintained at a relative 
phase difference $\phi$ across the metallic region (see Fig.~\ref{fig:SNS_junction}). Such a phase 
difference is known to give rise to Andreev bound states 
(ABS)~\cite{Satoshi_Kashiwaya_2000}, which appear as subgap features in the 
local density of states and whose energies disperse with $\phi$. Previous studies have focused on a variety of domain-wall configurations that involve chiral or otherwise 
unconventional order parameters, revealing characteristic ABS dispersions 
~\cite{mukherjee2015domain, samokhin2012domain, Satoshi_Kashiwaya_2000}, here we extend this analysis to explore how domains whose order parameters differ by an overall 
phase $\phi$ influence the attenuation coefficient $\alpha_\mathrm{s}$. Phenomenological models have suggested that 
external perturbations such as acoustic waves may affect domain-wall structures~\cite{Revsigrist}. 
Here, we build upon these insights by providing a fully microscopic calculation of the attenuation coefficient 
$\alpha_\mathrm{s}$ for L100 mode in such heterostructures.
\\
In our calculations, we employ a tight-binding model and onsite superconducting pairing interactions that generate 
an $s$-wave superconducting order parameter. The system considered here has a ribbon configuration having periodicity along the $y$-axis. It consists of a one-dimensional chain comprising 
$351$ lattice sites along the $x$-axis: $175$ sites on the left and $175$ sites on the right, separated by
a short normal-region junction of a single site (with $d = 1$ and $l = 175$), as illustrated in 
Fig.~\ref{fig:SNS_junction}.
The superconducting order parameter has conventional $s$-wave symmetry that corresponds to an onsite pairing in real space ($\Delta_{ij}(k_y)=\Delta_i \delta_{ij}$). 
For the SNS junction geometry shown in Fig.~\ref{fig:SNS_junction}, the Hamiltonian given in Eq.~\eqref{eq:benchmarkHamiltonian} consists of left (right) superconductor, with the superconducting gap magnitude calculated self-consistently but the phase is kept fixed at $e^{-i\phi/2}$ ($e^{i\phi/2}$) respectively. Additionally, the pairing interaction $\text{V}=0$ is maintained at the normal metal interface. The electronic structure and pairing strength are identical to those used for the $s$-wave case in Sec.~\ref{sec:2}. The temperature is fixed at $T=0.001$.

In Fig.~\ref{fig:DOS_ABS}(a), we show the variation of density of states (DOS) with phase difference 
$\phi$ across the junction. The phase bias induces the formation of subgap Andreev bound states (ABS) within the superconducting 
gap, whose energies depend on $\phi$. In the short ballistic limit as in our case,  the bound-state energy follows $ E_{\rm bs}(\phi)
    =
    \pm \Delta_0 \cos{\phi /}{2}.$ 
 Since ABS occur in time-reversed pairs, two subgap peaks 
appear symmetrically in the DOS, as illustrated in Fig.~\ref{fig:DOS_ABS}(a). For $\phi$ near $\pm \pi$, the ABS peak lies at zero 
energy; as $\phi$ decreases toward zero, the two states split in energy and move toward the coherence peak at $\pm \Delta_0 $. 
At $\phi = 0$, the ABS merges with the coherence peak. In Fig.~\ref{fig:DOS_ABS}(b), we present the local density of states (LDOS) for $\phi=0.6\pi$, showing the spatial extent of ABS across the junction, having coherence length $\xi_\mathrm{c} \sim 9 a$. 

The attenuation coefficient ratio $\alpha_\mathrm{s} / \alpha_\mathrm{n} $ is calculated using the real-space methodology described in Sect.~\ref{sec:1}. Fig.~\ref{fig:SNS}(a) shows the attenuation for $\omega_q \approx 0.009$ (hereafter, we set $\hbar=1$) as a function of the phase 
difference $\phi$, exhibiting a pronounced peak near $\phi = \pm \pi$. This behavior can be attributed to 
the splitting of Andreev bound states (ABS), $\Delta E_{\text{bs}} \approx \omega_q$. The
electron--phonon scattering cross-section  increases for frequencies $\omega_q $ comparable to splitting energy $\Delta E_{\text{bs}}$ (within the broadening factor) and thus produces a peak in attenuation. As $\omega_q$ increases, this 
peak gradually shifts away from $\phi = \pi$, as seen in Fig.~\ref{fig:SNS}(b).

To further investigate this trend, we compute the ratio $\alpha_\mathrm{s} / \alpha_\mathrm{n} $ for several phonon frequencies 
$\omega_q$, as shown in Fig.~\ref{fig:SNS}(c). The results indicate that the peak attenuation $\alpha_{\max}$ occurs at a distinct 
phase value $\phi$ for each $\omega_q$, and that these peak positions follow an approximately linear relationship near $\phi = \pi$. 
The phase values at which $\alpha_{\max}$ (red dots) occurs are compared with the corresponding ABS energy 
splittings $\Delta E_{bs}$ (black dots). Interestingly, the energy difference of the bound-states $\Delta E_{bs}$ closely matches $\omega_q$ for each $\phi$, indicating that the resonance condition $\Delta E_{\text{bs}}(\phi) \approx \omega_q$ is fulfilled.
Whenever this resonance condition is satisfied, an enhancement in the electron--phonon scattering cross-section 
is expected, giving rise to a peak in the attenuation coefficient. 

Although conceptually reminiscent of Josephson 
junction experiments, this mechanism operates at a much lower energy scale of phonons, which is several orders of 
magnitude smaller than typical electronic energies. It offers a potential approach for detecting 
large phase differences (or small bound state energies) in SNS junctions.
\section{Attenuation near Grain Boundary in d-wave superconductors}
\label{sec:4}
\begin{figure}[t]
 \includegraphics[width=\linewidth]{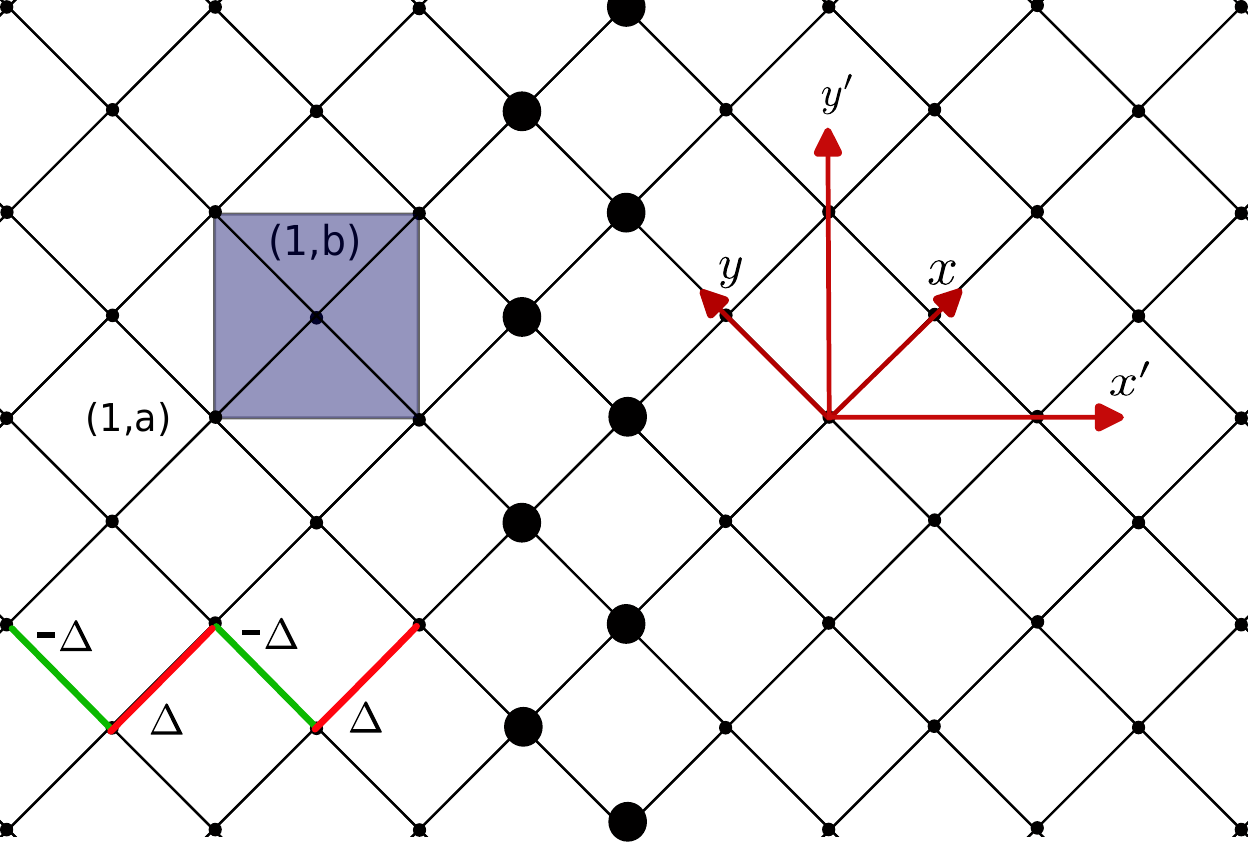}
\caption{
The lattice considered here is a rotated square lattice, obtained by transforming the basis from $(x,y)$ to $(x',y')$. The shaded region denotes the two-sublattice unit cell (a and b). Two impurity lines are introduced, one on each sublattice (solid black dots). The superconducting order parameter $\Delta$ shown corresponds to the $d$-wave symmetry.}
  \label{fig:lattice_lineimpurity}
\end{figure}

As a second example of the study of ultrasonic attenuation for inhomogeneous superconductors, we simulate a grain boundary between two superconducting regions by two neighboring impurity lines oriented along the (110) direction on a square lattice. The system is therefore translationally invariant along the direction of the grain boundary but breaks translational symmetry along the transverse direction.  
To exploit this symmetry, we rotate the basis $(x,y)$ by $45^\circ$ to $(x',y')$ and define two sublattices (a and b), as shown in Fig.~\ref{fig:lattice_lineimpurity}. 
We consider that the superconducting gap function has a $d_{x^2-y^2}$ symmetry (or $d_{x'y'}$ on a rotated basis) which corresponds to superconducting gaps stabilized on nearest-neighbor bonds in real space. The (110) grain boundary has been modeled with 2 impurity lines having a non-magnetic impurity potential $\text{V}_\mathrm{imp}$, as shown in Fig.~\ref{fig:lattice_lineimpurity}. For a $d_{x'y'}$ superconducting gap function, a (110) grain boundary gives rise to localized bound states~\cite{PhysRevLett.72.1526}. These bound states originate from specular reflection of quasiparticles from the positive to negative superconducting phase lobes at the junction regions. Due to the large degeneracy associated with momentum parallel to the boundary, these bound states have considerable spectral weight. The splitting of the bound states depends sensitively on the impurity strength. As the impurity strength increases, the bound states continuously shift from the gap edges toward zero energy. Note that for a (100) grain boundary, no bound state is expected to form, as the reflection of quasiparticles from such a junction connects the superconducting phases with the same sign. Significant work has been carried out on Andreev bound states (ABS) across insulator junctions, and several review articles summarize these developments~\cite{lofwander2001andreev,tanaka1995theory, balatsky2006impurity}.
For the grain boundary problem, the Hamiltonian can be expressed in terms of the conserved Bloch momentum $k_y$ and is given by
\begin{align}
\mathcal{H}(k_y)
&= \sum_{\substack{i,j,\sigma \\ \alpha,\beta}}
t_{ij}^{\alpha\beta}(k_y)\,
c^{\dagger}_{i\alpha\sigma}(k_y)\,
c_{j\beta\sigma}(k_y) \notag\\
&\quad  +\sum_{i,\alpha,\sigma} \left(V_{\mathrm{imp}}\delta_{i,i_\mathrm{imp}}- \mu\right)
c^{\dagger}_{i\alpha\sigma}(k_y)\,
c_{i\alpha\sigma}(k_y) \notag  \\
&\quad + \sum_{i,j,\alpha,\beta}
\left[
\Delta_{ij}^{\alpha\beta}(k_y)\,
c^{\dagger}_{i\alpha\uparrow}(k_y)\,
c^{\dagger}_{j\beta\downarrow}(-k_y)
+ \text{H.c.}
\right].
\label{eq:Hamiltonian_h0_LineImpurity}
\end{align}
Here, $i$ and $j$ label the coordinates along the $x$ direction, while
$\alpha,\beta \in \{a,b\}$ denote the sublattice indices. The parameters $t_{ij}^{\alpha\beta}(k_y)$ and $\mu$ represent the hopping amplitudes and the chemical potential, respectively. The parameter $V_{\mathrm{imp}}$ denotes the strength of the line impurity located at the sites $i=i_\mathrm{imp}$. The system is considered periodic in the $y$ direction. The pairing amplitude $\Delta_{ij}^{\alpha\beta}(k_y)$ encodes the $d_{x^2-y^2}$ superconducting order parameter.
To solve the Hamiltonian in Eq.~\eqref{eq:Hamiltonian_h0_LineImpurity}, we utilize a Bogoliubov–de Gennes (BdG) transformation, 
 \begin{align}  
{c}^\dagger_{{i} \alpha \sigma}(k_y)=\sum_{n >0 } {u_{{i} \alpha \sigma}^{n *}}^{}(k_y){\gamma}^\dagger_{{n} ,k_y }-{\sigma}v_{{i} \alpha \sigma}^{n }(k_y){\gamma}_{n ,-k_y}
\label{eq:bdgsublatticetransformation}
\end{align}
The superconducting pairing amplitude is obtained self-consistently from the BdG eigenfunctions as 
\begin{align}
&\Delta_{ij}^{\alpha\beta}(k_y)
=
\frac{V M_{ij}^{\alpha\beta}(-k_y)}{2 N_{y}}\,
\sum_{n,k_y'}  \left( M_{ij}^{\alpha\beta}(k_y')  u_{i \alpha \uparrow}^n(k_y') v_{j \beta \downarrow}^{n *}(-k_y') \right. \notag \\&  
 f(-E_n)- \left. M_{ji}^{\beta\alpha}(k_y')u_{j \beta \uparrow}^n(k_y') v_{i \alpha \downarrow}^{n *}(-k_y')f(E_n)\right) 
\label{eq:gap_equation}
\end{align}
where $V$ denotes the strength of the pairing interaction, 
$u_{i \alpha \sigma}^n$ and $v_{i \alpha \sigma}^n$ are eigencomponents of the $n$-th BdG eigenstate with energy $E_n$ corresponding to the site $i$, sublattice $\alpha$ and spin $\sigma$. The pairing form factor
$M_{ij}^{\alpha\beta}(k_y)$ reflects the symmetry $d_{x'y'}$ of the order parameter, written as
\begin{align}
M_{i j}^{\alpha \beta}\left(k_y\right)=& 2 \mathrm{i} \sin \left(\frac{k_y}{2}\right)\left( \delta_{\alpha a} \delta_{\beta b}\left(\delta_{i j}-\delta_{i-1, j}\right) \right.\notag \\
& \left.+\delta_{\alpha b} \delta_{\beta a}\left(\delta_{i+1, j}-\delta_{i j}\right)\right) .
\label{eq:form_factor}
\end{align}
As we have considered the inter-sublattice separation in the Bloch phase factors, the conserved momentum $k_y$ is defined over an extended Brillouin zone, $k_y \in [0, 4\pi)$. 
Near grain boundaries created by impurity potentials in unconventional superconductors, in-gap bound states are expected to occur at finite energies. In unitary limit, these in-gap bound states start approaching zero energy but can be spin split in energy (with a splitting of less than 0.002) if electron correlations generate intrinsic disorder induced magnetism (DIM)~\cite{matsubara2020emergence,PhysRevB.110.064502}. In the following, we compare the scenarios for the grain boundary formed by impurity without DIM and strong impurity with DIM (see Fig.~\ref{fig:Schematic_Spinsplitting}). We show that both scenarios can generate particle-hole symmetric bound state peaks at similar energy scales that cannot be distinguished in non spin polarized STM experiments. Next, we ask whether ultrasonic attenuation experiments could distinguish these two scenarios.
\begin{figure}[]
    \centering
        \includegraphics[width=\linewidth]{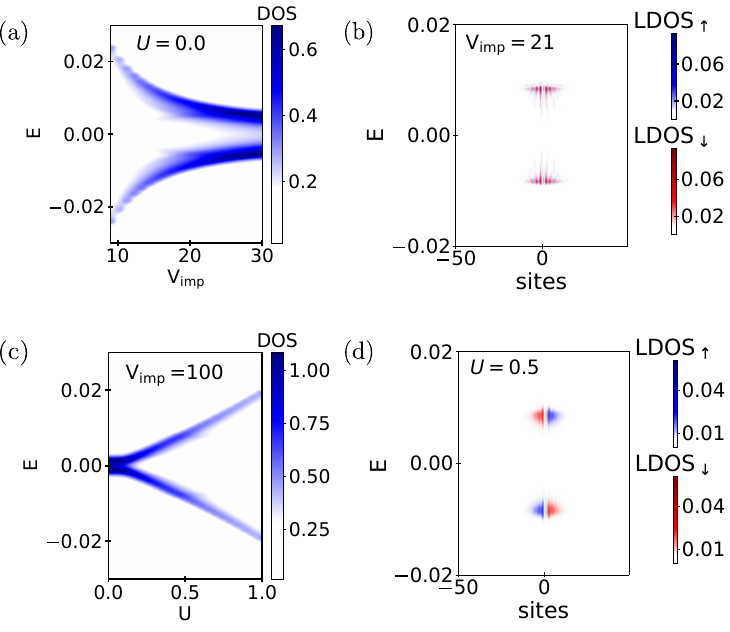} 
        \caption
{Plots for the non-magnetic bound state and magnetic bound state cases.
(a) Density of states, $N(E)$, as a function of the impurity strength $V_{\mathrm{imp}}$ for the non-magnetic bound state.
(b) Spin-resolved local density of states for the non-magnetic bound state. Red and blue denote the spin-down and spin-up components, respectively, while purple indicates an equal admixture of the two spin components. The local density of states is obtained from the imaginary part of the normal Green's function.
(c) Density of states, $N(E)$, as a function of the onsite Coulomb interaction strength $U$ for the magnetic bound state.
(d) Spin-resolved local density of states for the magnetic bound state.
} 
\label{fig:Dos_Vimp}
\end{figure}
\subsubsection{Non-magnetic bound state}
First, we will analyze the impact of impurity strength on the attenuation coefficient using the Hamiltonian, Eq.\eqref{eq:Hamiltonian_h0_LineImpurity}. 
The electronic structure and pairing strength are identical to those used for the $d$-wave case in Sec.~\ref{sec:2}. The temperature is fixed at $T=0.001$. 
\begin{figure}[t]
    \centering
        \includegraphics[width=\linewidth]{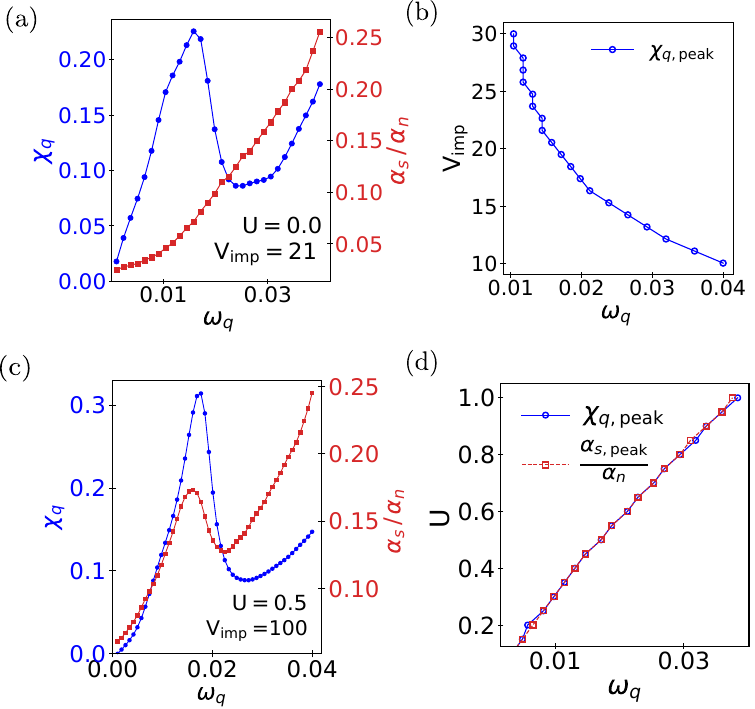}       
\caption{
{Susceptibility function ${\chi}_{\mathbf{q}}$ and attenuation ratio $\alpha_{\mathrm{s}}/\alpha_{\mathrm{n}}$ for the non-magnetic bound state  and magnetic bound state case.}
(a) $\chi_{\mathbf{q}}$ and $\alpha_{\mathrm{s}}/\alpha_{\mathrm{n}}$ as functions of $\omega_{\mathbf{q}}$ for the non-magnetic bound state.
(b) $\omega_{\mathbf{q}}$ plotted against $V_{\mathrm{imp}}$ at maximum values of $\chi_{\mathbf{q}}$.
(c) $\chi_{\mathbf{q}}$ and $\alpha_{\mathrm{s}}/\alpha_{\mathrm{n}}$ as functions of $\omega_{\mathbf{q}}$ for the magnetic bound state with $U=0.5$.
(d) $\omega_{\mathbf{q}}$ plotted against $U$ at maximum values of $\chi_{\mathbf{q}}$ and $\alpha_{\mathrm{s}}/\alpha_{\mathrm{n}}$ for magnetic bound states.
}
\label{fig:Chi_alpha_LineImpurity_Vimp}
\end{figure}

For a specific range of impurity potentials, we see splitting in ABS as shown in Fig.~\ref{fig:Dos_Vimp}(a). This splitting (as in previous subsection) leads to the enhancement in electron-phonon scattering cross-section and can be understood better from the susceptibility function ${\chi}_{\mathbf{q}} $ (see Eq.~\eqref{eq:Lindhard_function}). This function shows the scattering cross-sections corresponding to the processes involving scattering between the split bound-state as peaks, in addition to the contributions from background scattering of Bogoliubov-de Gennes (BdG) quasiparticles. In Fig.~\ref{fig:Chi_alpha_LineImpurity_Vimp}(a), we present the variation of ${\chi}_{\mathbf{q}} $ as a function of the sound frequency for the impurity strength $V_{\mathrm{imp}} \approx 21\,$ (blue curve).
The peaks in ${\chi}_{\mathbf{q}} $ occur at frequencies that match the energy difference between split ABS, but interestingly as shown in Fig.~\ref{fig:Chi_alpha_LineImpurity_Vimp}(a),(b) (red curve) these features do not translate to the attenuation coefficient ratio $\alpha_{\mathrm{s}}/\alpha_{\mathrm{n}}$ at similar energies.

The absence of a clear bound-state feature in attenuation should not be interpreted as the absence of resonant in-gap states. The split bound states do appear in the susceptibility function when the phonon energy matches their separation, $\omega_q \simeq |E_m-E_n|$. However, the attenuation coefficient contains not only this resonant denominator, but also the electron--phonon coherence factor $\mathcal{F}^\lambda_{mn}(q)$. For a non-magnetic grain boundary, the two split bound states are related by particle--hole symmetry. If $n=\bar m$ denotes the particle--hole partner of the state $m$, in the presence of a time reversal symmetry the eigenstates and eigenvalues at $n$ and $\bar m$ correspond to two separate blocks of the BdG Hamiltonian, and the corresponding BdG amplitudes satisfy, up to the usual spin-dependent phase convention,
\[
u^{\bar m}_{i\sigma}=-\sigma v^{m*}_{i\sigma},
\qquad
v^{\bar m}_{i\sigma}=-\sigma u^{m*}_{i\sigma}.
\]
Substituting these relations into the coherence factor shows that the electron-like and hole-like parts of the matrix element cancel for the particle--hole-conjugate pair, yielding $\mathcal{F}^\lambda_{m\bar m}(q)=0 $. As a result, the direct transition between the two particle--hole-symmetric bound states does not contribute to attenuation: the states are visible in the DOS and in the susceptibility function, but they do not produce a corresponding peak in the ultrasonic attenuation, see Fig.~\ref{fig:Chi_alpha_LineImpurity_Vimp}(a). The remaining finite background in $\alpha_{\mathrm{s}}$ comes from other quasiparticle processes for which $E_m\neq -E_n$. Even though the particle hole symmetry of the BdG Hamiltonian remains intact, this selection rule can be lifted when the symmetry structure of the bound states is modified, for example by a phase-textured SNS junction or by local magnetization near the grain boundary, in which case the bound-state resonance can acquire finite coherence factors.
\begin{figure}[t]
    \centering
%     \subfloat[]{%
        \includegraphics[width=\linewidth]{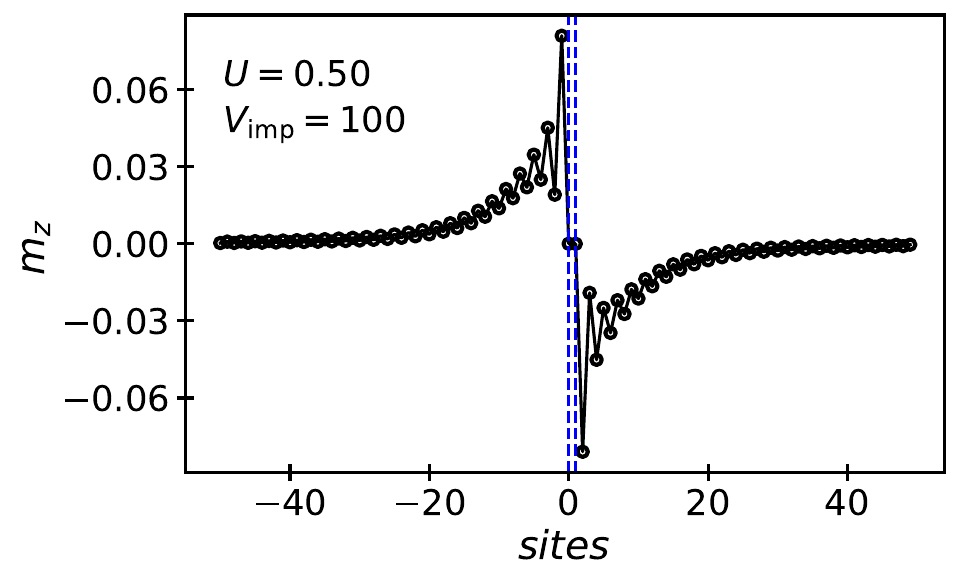}
%     }
% 
\caption{Magnetic bound state: Local magnetic moments are induced across the impurity line even for onsite interaction strengths $U$ below the critical Stoner threshold $U_c$. 
Results are shown for $U = 0.5$.}
\label{fig:SpatialMoments_lineimpurity}
\end{figure}
\subsubsection{Magnetic bound states }
In the following, we study the unitary limit impurity regime, choosing $V_{\mathrm{imp}} \approx 100$, which leads to the formation of a bound state at zero energy. The system has the same electronic structure and superconducting gap as in the previous case. The critical Stoner interaction is $U_c \approx 5$, which corresponds to a density of states at the Fermi level of $N(E_F) \approx 0.2$. Additionally, we introduce the onsite Coulomb interaction described by a Hubbard Hamiltonian, which at the mean-field level is given by 
\begin{equation}
H_{\mathrm{U}}
=
U \sum_{i,\alpha,\sigma, k_y}
\left\langle n_{i\alpha\bar{\sigma}} \right\rangle
\, c^{\dagger}_{i\alpha\sigma}(k_y) c_{i\alpha\sigma}(k_y)
\end{equation}
where $\langle n_{i\alpha\sigma} \rangle = \frac{1}{N_y}\sum_{k_y} c^{\dagger}_{i\alpha\sigma}(k_y)\, c_{i\alpha\sigma}(k_y)$ is the average electron density at site $i$, sublattice $\alpha$, and spin $\sigma$. Here, $\bar{\sigma}$ denotes the spin opposite to $\sigma$. The self-consistent solution leads to the formation of a local DIM due to the local crossing of Stoner instability near the grain boundary region.
As the onsite interaction $U$ increases, local magnetic moments are induced and become more pronounced near the impurity line, as illustrated in Fig.~\ref{fig:SpatialMoments_lineimpurity}. Concurrently, the zero-energy bound states localized near the impurity line undergo a splitting, as evident from the density of states (DOS) shown in Fig.~\ref{fig:Dos_Vimp}(c). These bound states have spin-dependent spectral features across the junction. It can be understood well by the spin-resolved single-particle Green's functions, whose imaginary parts, $\mathrm{Im}\,G_{\uparrow\uparrow}(E)$ and $\mathrm{Im}\,G_{\downarrow\downarrow}(E)$, correspond to the local density of states (LDOS) for each spin.
We therefore find that both the non-magnetic bound state case without any local DIM (see Fig.~\ref{fig:Dos_Vimp}(b)) and the magnetic bound state case with DIM (see Fig.~\ref{fig:Dos_Vimp}(d)) give rise to similar particle--hole symmetric in-gap bound states. However, only in the latter case are the bound states spin selective.
These bound-state splittings produce a resonant peak in
$\chi_{\mathbf{q}}$, as shown by the blue curve in
Fig.~\ref{fig:Chi_alpha_LineImpurity_Vimp}(c). The resonance
occurs when the phonon frequency matches the energy difference
between the relevant bound states,
$\omega_{\mathbf{q}} \simeq \lvert E_m-E_n\rvert$. In contrast
to the nonmagnetic bound-state case, the corresponding
resonance is also visible in the attenuation ratio
$\alpha_{\mathrm{s}}/\alpha_{\mathrm{n}}$, as shown by the red
curve in Fig.~\ref{fig:Chi_alpha_LineImpurity_Vimp}(c).
Figure~\ref{fig:Chi_alpha_LineImpurity_Vimp}(d) shows that both
the susceptibility function and attenuation peaks shift to higher phonon
frequencies with increasing $U$. This behavior results from the
enhancement of the local magnetic moment near the grain boundary,
which increases the splitting between the positive- and
negative-energy in-gap bound-state branches.

The difference between the magnetic and nonmagnetic cases
originates from the coherence factors weighted electron--phonon matrix elements $\mathcal F^{\lambda}_{mn}(\mathbf{q})$ in
Eq.~\eqref{eq:coherencefactors}. Although
$\chi_{\mathbf{q}}$ identifies energetically allowed transitions,
their contribution to the attenuation is weighted by
$\lvert \mathcal F^{\lambda}_{mn}(\mathbf{q})\rvert^2$. Since the
electron--phonon interaction considered here conserves spin, only
components of the two BdG states with compatible spin character
contribute to this matrix element. On a given side of the grain
boundary, the positive- and negative-energy magnetic bound states
predominantly have opposite spin character, and the corresponding
same-side contributions to
$ \mathcal F^{\lambda}_{mn}(\mathbf{q})$ are therefore strongly suppressed.
However, the spin polarization of the bound states reverses across
the grain boundary, as shown in Fig.~\ref{fig:Dos_Vimp}(d).
Consequently, terms for which the site indices $i$ and $j$ lie on
opposite sides of the boundary can connect components with the same
physical spin and give a finite contribution to the coherence
factor.

Based on the symmetry of the BdG Hamiltonian in the Nambu basis, the above arguments can be summarized in the following manner. 
In the nonmagnetic problem, time-reversal symmetry relates the two spin–Nambu blocks of the BdG Hamiltonian. 
Together with BdG particle–hole symmetry, this produces an effective symmetry within either reduced block that relates the positive- and negative-energy bound-state wavefunctions. 
For the spin-conserving electron–phonon vertex, the electron-like and hole-like contributions to $\mathcal{F}^\lambda_{m\bar m}(q)$ then cancel exactly. Local magnetism breaks time-reversal symmetry and makes the two spin–Nambu blocks inequivalent. 
Although the full BdG Hamiltonian retains particle–hole symmetry, the intrablock wavefunction relation responsible for the cancellation no longer applies. The corresponding electron–phonon matrix element is therefore no longer symmetry-forbidden and is generically finite.

\section{Conclusions}
We have developed a real space model to study ultrasonic attenuation in unconventional superconductors and show that it can be a useful probe for identifying unconventional superconducting order, in-gap bound state physics and local magnetism near inhomogeneities. For SNS junctions in which the two s-wave superconductors have a relative phase difference leading to formation of in-gap bound states near the junction region, ultrasonic attenuation can be used to detect these bound states with a peak in attenuation coefficient forming when the phonon energy matches with the energy separation of the particle–hole symmetric bound states. For the case of grain boundary in a $d$-wave superconductor, similar hump-like features in attenuation coefficient will not be observed at the bound state energies unless a physical mechanism exists that breaks the symmetry of the BdG Hamiltonian. We verify this by showing that the characteristic features in $\alpha_\mathrm{s}/\alpha_\mathrm{n}$ near the bound state energy are indeed observed if electron correlations lead to local breaking of time-reversal symmetry near grain boundaries. This local symmetry breaking results in in-gap bound states being spin selective. In general, the peak in ultrasonic attenuation is obtained when the resonance condition ($\omega_q=2E_B$) is satisfied in the charge susceptibility function, the attenuation is also intricately dependent on coherence factors. For inhomogeneous superconductors, these coherence factors become site selective and can help filter local observables near inhomogeneities even though the measurement remains nonlocal.
\begin{figure}[t]
    \centering
        \includegraphics[width=\linewidth]{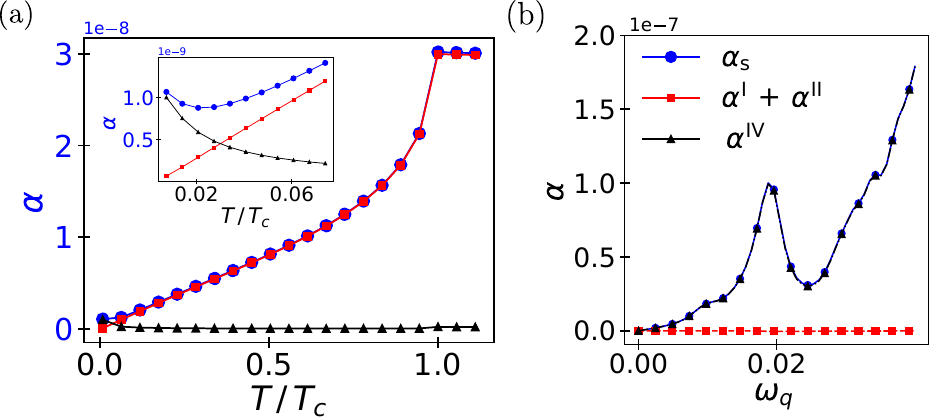} 
\caption{Contribution to the ultrasonic attenuation from phonon-induced pair creation and annihilation of two BdG quasiparticles, $\alpha_{\mathrm{s}}^{\mathrm{IV}}$. (a) For the $d$-wave $L110$ mode and $\omega_q\approx 0.005$, the pair creation process from $\alpha_{\mathrm{s}}^{\mathrm{IV}}$ dominates for $T<<T_c$. For comparison, the total attenuation (blue) and $\alpha_{\mathrm{s}}^{\mathrm{I}}+\alpha_{\mathrm{s}}^{\mathrm{II}}$ (red) are also shown. (b) The presence of magnetic bound states (strong impurity) enhances the $\alpha_{\mathrm{s}}^{\mathrm{IV}}$  contribution to the attenuation at low temperatures.
}
\label{fig:QPSIII_IV}
\end{figure}
In terms of the individual electron-phonon interaction processes, we show in Fig.~\ref{fig:QPSIII_IV}(b) that these attenuation features at the bound state energies are contributed primarily by processes involving phonon creation or annihilation from two BdG quasiparticles ($\alpha_{\mathrm{s}}^{\mathrm{IV}}$ term in Appendix~\ref{app:1}) as this process dominates the attenuation coefficient at low temperatures compared to the superconducting transition temperature. At higher temperatures, the attenuation is dominated by phonon scattering processes (see Fig.~\ref{fig:QPSIII_IV}(a)) that combined with usual thermal broadening effects would wash out the signatures for in-gap bound states in the ultrasonic attenuation.

These results indicate that a combination of STM and attenuation measurements could provide an indirect probe of the spin structure of bound states and the presence of local magnetism in unconventional superconductors. The form of a local magnetic instability near grain boundaries, step edges, or surfaces will depend on details of the electronic structure, disorder profile, and superconducting gap symmetry. The present framework can therefore provide a basis for material-specific predictions of ultrasonic attenuation near spatial inhomogeneities. In the calculations presented here, the chosen bound-state splittings correspond to phonon frequencies in the terahertz range. This choice allows the relevant resonances to be resolved using computationally accessible system sizes, momentum grids, and broadening parameters. The underlying resonance mechanism is expected to persist at lower frequencies, although a quantitative demonstration at such scales would require correspondingly smaller numerical broadenings and finite-size level spacings. 

Although conventional ultrasonic attenuation experiments typically operate in the MHz-to-GHz range, recent developments in picosecond ultrasonics and terahertz phononic spectroscopy have enabled the generation and detection of coherent acoustic phonons from tens of gigahertz to several terahertz~\cite{yoon2024terahertz,zalalutdinov2021acoustic,pupeikis2021picosecond}. These approaches could potentially extend bound-state spectroscopy to the larger energy splittings considered here. The same framework can be applied to topological edge and surface states to determine whether their zero-energy spectral weight is acoustically active or suppressed by particle–hole coherence factors. The results of this work could also allow future material-specific studies of superconducting state and surface state properties of unconventional superconductors such as UTe$_2$ and Sr$_2$RuO$_4$. In these materials, ultrasonic attenuation experiments have been interpreted as having possible evidence of unusual superconducting gap symmetry, domain wall physics, time reversal symmetry breaking, or surface superconducting states ~\cite{doi:10.1126/science.adk7219,PhysRevLett.107.077003,csire2026directionalandreevreflectionsignaturesinterorbital,wang2025odd,christiansen2025nodal, ando2022surface}.
\begin{acknowledgments}
A.~K. acknowledges support by the Danish National Committee for Research Infrastructure (NUFI) through the ESS-Lighthouse Q-MAT.
A.K. acknowledges support by the
Institute of Eminence (IoE) program of IIT Madras.
\end{acknowledgments}

\appendix

\section{\text{$\alpha_{\mathrm{s}}\left({\mathbf{q},\mathbf{\lambda}}\right) $} for Inhomogeneous Superconductors}
\label{app:1}
\begin{figure}[t]
        \includegraphics[width=\linewidth]{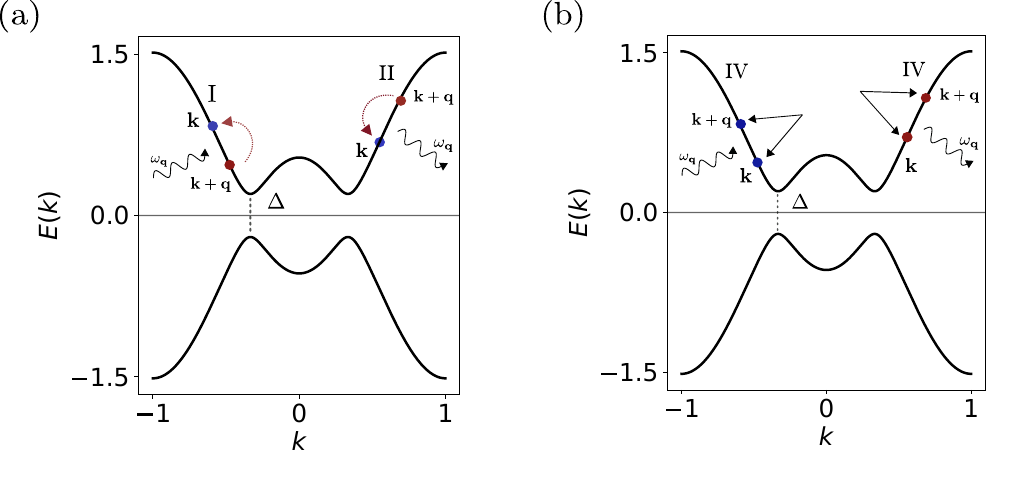} 
\caption{ Schematic illustration of the scattering processes contributing to the attenuation coefficient $\alpha_{\mathrm{s}}\left({\mathbf{q},\mathbf{\lambda}}\right)$. (a) PH channel: Processes I and II describe single-quasiparticle scattering accompanied by phonon absorption and emission, respectively. (b) PP channel: Processes IV correspond to pair creation and pair annihilation of BdG quasiparticles accompanied by phonon annihilation and creation, respectively. Blue (red) circles denote quasiparticle creation (annihilation). The quasiparticle energies are labeled as $E_m = E_{\mathbf{k}+\mathbf{q}}$ and $E_n = E_{\mathbf{k}}$ (Eqs.~\eqref{eq:alpha_1}--\eqref{eq:alpha_4}).
}
\label{fig:dispersion_scattering_BQP}
\end{figure}

In this Appendix, we evaluate the attenuation coefficient $\alpha_{\mathrm{s}}\left({\mathbf{q},\mathbf{\lambda}}\right)$ in the superconducting phase. The calculation is carried out by expressing the commutator in Eq.~\eqref{eqn:alphaSCDraft} in the Bogoliubov--de Gennes (BdG) quasiparticle basis Eq.~\eqref{eq:bdgtransformation}. This commutator can be categorized into two types of scattering processes: BDG quasiparticle-quasihole (PH) and BdG quasiparticle-quasiparticle (PP) scattering. These two channels are discussed separately below. We first focus on the PH contribution.

\begin{align}
&\left[ c_{\mathbf{i}\sigma}^{\dagger}(t) c_{\mathbf{j}\sigma}(t),c^\dagger_{\mathbf{i}^\prime\sigma^\prime} c_{\mathbf{j}^\prime\sigma^\prime}\right]_{\mathrm{PH}}= \sum_{\substack{m,n>0 \\ {m',n' >0}} } \left[ {u_{\mathbf{i} \sigma}^{m *}}^{} {u_{\mathbf{j} \sigma}^{n }}^{}{\gamma}^\dagger_{{m} }(t){\gamma}_{n}(t) \right. \notag\\
    &\left. +v_{\mathbf{i} \sigma}^{m }v_{\mathbf{j} \sigma}^{n * }{\gamma}_{m}(t){\gamma}_{n}^\dagger(t) , \hspace{0.05 cm}{u_{\mathbf{i'} \sigma'}^{m' *}}^{} {u_{\mathbf{j'} \sigma'}^{n' }}^{}{\gamma}^\dagger_{{m'} }{\gamma}_{n'}
+v_{\mathbf{i'} \sigma'}^{m' }v_{\mathbf{j'} \sigma'}^{n' * }{\gamma}_{m'}{\gamma}_{n'}^\dagger\right] 
\label{eq:ScCommutatorbDgPH}
\end{align}
The above expression consists of terms of the form $\gamma^\dagger \gamma$ and $\gamma \gamma^\dagger$, which describe single-quasiparticle scattering processes mediated by phonon absorption or emission, as shown in Fig.~\ref{fig:dispersion_scattering_BQP}(a). Upon evolving the operators $\gamma_m^\dagger(t)$ and $\gamma_m(t)$ under the superconducting Hamiltonian $\mathcal{H}_{\mathrm{sc}}$ (Eq.~\eqref{eq:base_sc_ham}) and applying Wick's theorem, Eq.~\eqref{eqn:alphaSCDraft} can be written as

 \begin{align}
&\alpha_{\mathrm{s}}\left({\mathbf{q},\mathbf{\lambda}}\right)= \frac{\ell_{\mathbf{q}\mathbf{\lambda}}^2}{{2\hbar^2 v_s N}}  \sum_{m,n >0}\sum_{\substack{\mathbf{ij} \sigma\\\mathbf{\mathbf{i}^\prime \mathbf{j}^\prime} \sigma'}} \int_{0}^{\infty} {d} t  { \delta t_{\mathbf{ij}}^\lambda\delta t_{\mathbf{i}^\prime \mathbf{j}^\prime}^\lambda }\chi^{mn}_{{\mathbf{ij}} {\mathbf{\mathbf{i}^\prime \mathbf{j}^\prime}}}(\sigma,\sigma')  \notag \\ 
& \times (f_m-f_n) \mathrm{e}^{i\left(E_m-E_n \right)t / \hbar}  \left( \beta_{\mathbf{ij} \mathbf{i^\prime j^\prime}}  \left(\mathbf{q}\right) e^{i\omega_\mathbf{q} t}-\beta_{\mathbf{ij} \mathbf{i^\prime j^\prime}}^*  \left(\mathbf{q}\right)  e^{-i\omega_\mathbf{q} t}\right)
\label{eq:A2}
\end{align}
where the coherence factors $\chi^{mn}_{{\mathbf{ij}}{\mathbf{i}^\prime \mathbf{j}^\prime}}( \sigma, \sigma^\prime)$ are
\begin{align}
       &\chi^{mn}_{{\mathbf{ij}} {\mathbf{i}^\prime \mathbf{j}^\prime}}  ( \sigma, \sigma^\prime)= \ 
      {u_{\mathbf{i} \sigma}^{m *}} {u_{\mathbf{j} \sigma}^{n }} {u_{\mathbf{i}^\prime {\sigma'}}^{n *}} {u_{\mathbf{j}^\prime {\sigma'}}^{m }}  - u_{\mathbf{i} \sigma}^{m *} {u_{\mathbf{j} \sigma}^{n }} v_{\mathbf{i}^\prime {\sigma'}}^{m } v_{\mathbf{j}^\prime {\sigma'}}^{n  *}   \notag \\& \qquad \qquad
     + {v_{\mathbf{i} \sigma}^{n }} {v_{\mathbf{j}\sigma}^{m *}} {v_{\mathbf{i}^\prime {\sigma'}}^{m }} {v_{\mathbf{j}^\prime {\sigma'}}^{n *}} 
      - v_{\mathbf{i} \sigma}^{n } v_{\mathbf{j} \sigma}^{m *} u_{\mathbf{i}^\prime {\sigma'}}^{n *} {u_{\mathbf{j}^\prime {\sigma'}}^{m }}.
\end{align}

Using the Sokhotski--Plemelj theorem, the time integral in Eq.~\eqref{eq:A2} can be evaluated straightforwardly. The resulting expression contains contributions from the two scattering processes labeled as I and II in Fig.~\ref{fig:dispersion_scattering_BQP}(a). We write these corresponding terms separately as
\begin{align}
    & \alpha_{\mathrm{s}}^{\mathrm{I}}\left({\mathbf{q},\mathbf{\lambda}}\right)=    \frac{\pi  \ell_{\mathbf{q}\mathbf{\lambda}}^2}{{2\hbar v_s N}} \sum_{m,n>0} \sum_{\substack{\mathbf{ij}  \\ \mathbf{\mathbf{i}^\prime \mathbf{j}^\prime}}}  {{ \delta t_{\mathbf{ij}}^\lambda\delta t_{\mathbf{i}^\prime \mathbf{j}^\prime}^\lambda }\chi_{\mathbf{i j i^\prime j^\prime}}^{mn}} (\sigma ,\sigma')    \notag \\& \quad \qquad\times (f_m-f_n) \beta_{\mathbf{ij} \mathbf{i^\prime j^\prime}} \left(\mathbf{q}\right)  \delta(E_m-E_n +\hbar \omega_\mathbf{q})
\label{eq:alpha_1}\end{align}
and 
\begin{align}
   &  \alpha_{\mathrm{s}}^{\mathrm{II}}\left({\mathbf{q},\mathbf{\lambda}}\right)  = - \frac{\pi  \ell_{\mathbf{q}\mathbf{\lambda}}^2}{{2\hbar v_s N}} \sum_{m,n>0} \sum_{\substack{\mathbf{ij}  \\ \mathbf{\mathbf{i}^\prime \mathbf{j}^\prime}}}    { \delta t_{\mathbf{ij}}^\lambda\delta t_{\mathbf{i}^\prime \mathbf{j}^\prime}^\lambda }\chi_{\mathbf{i j i^\prime j^\prime}}^{mn} (\sigma ,\sigma')   \notag \\& \quad \qquad \times (f_m-f_n) \beta_{\mathbf{ij} \mathbf{i^\prime j^\prime}}^* \left(\mathbf{q}\right) \delta(E_m-E_n -\hbar \omega_\mathbf{q})
\label{eq:alpha_2}\end{align}

This analysis can be done further for the PP channel in similar manner. In this case, the commutator in Eq.~\eqref{eqn:alphaSCDraft}, $\left[ c_{\mathbf{i}\sigma}^{\dagger}(t) c_{\mathbf{j}\sigma}(t),c^\dagger_{\mathbf{i}^\prime\sigma^\prime} c_{\mathbf{j}^\prime\sigma^\prime}\right]_{\mathrm{PP}}$ yields the following expansion,

\begin{align}
&\sum_{\substack{m,n>0 \\ {m',n' >0}} }\left[
    -{\sigma}{u_{\mathbf{i} \sigma}^{m *}}^{}v_{\mathbf{j} \sigma}^{n * }{\gamma}_{m}^\dagger(t) {\gamma}_{n}^\dagger(t) - {\sigma}v_{\mathbf{i} \sigma}^{m } {u_{\mathbf{j} \sigma}^{n }}^{}{\gamma}_{m}(t) {\gamma}_{{n} }(t) \hspace{0.1 cm} ,
     \right. \notag\\
    &\left.\qquad -{\sigma'}{u_{\mathbf{i'} \sigma'}^{m' *}}^{}v_{\mathbf{j'} \sigma'}^{n' * }{\gamma}_{m'}^\dagger{\gamma}_{n'}^\dagger  
    -{\sigma'}v_{\mathbf{i'} \sigma'}^{m' } {u_{\mathbf{j'} \sigma'}^{n' }}^{}{\gamma}_{m'} {\gamma}_{{n'} } \right] \notag
\end{align}
In contrast to the PH channel, the PP channel contains terms of the form $\gamma^\dagger \gamma^\dagger$ and $\gamma \gamma$. These terms describe processes involving the creation or annihilation of two BdG quasiparticles simultaneously, accompanied by the annihilation or creation of a phonon, respectively (see Fig.~\ref{fig:dispersion_scattering_BQP}(b)). Upon evolving the operators $\gamma_m^\dagger(t)$ and $\gamma_m(t)$ under the superconducting Hamiltonian $\mathcal{H}_{\mathrm{sc}}$ (Eq.~\eqref{eq:base_sc_ham}) and applying Wick's theorem for these processes, Eq.~\eqref{eqn:alphaSCDraft} can be again decomposed corresponding to two processes III and IV as,
    \begin{align}
        & \alpha_{\mathrm{s}}^{\mathrm{III}}\left({\mathbf{q},\mathbf{\lambda}}\right) = \frac{\pi  \ell_{\mathbf{q}\mathbf{\lambda}}^2}{{2\hbar v_s N}} \sum_{m,n>0} \sum_{\substack{\mathbf{ij}  \\ \mathbf{\mathbf{i}^\prime \mathbf{j}^\prime}}}    { \delta t_{\mathbf{ij}}^\lambda\delta t_{\mathbf{i}^\prime \mathbf{j}^\prime}^\lambda }  \delta\left(E_m+E_n + \hbar \omega_\mathbf{q} \right)\notag \\ & \times (1-f_m-f_n) \beta_{\mathbf{ij} \mathbf{i^\prime j^\prime}}\left(\mathbf{q}\right)  \left(  \xi ^{mn}_{{\mathbf{ij}} {\mathbf{\mathbf{i}^\prime \mathbf{j}^\prime}}} (\sigma,\sigma')+ \eta ^{mn}_{{\mathbf{\mathbf{i}^\prime \mathbf{j}^\prime}}\mathbf{ij}}  (\sigma,\sigma')\right) 
 \label{eq:alpha_3} 
 \end{align}
and 
% \begin{equation}
    \begin{align}
        & \alpha_{\mathrm{s}}^{\mathrm{IV}}\left({\mathbf{q},\mathbf{\lambda}}\right)  =  - \frac{\pi  \ell_{\mathbf{q}\mathbf{\lambda}}^2}{{2\hbar v_s N}} \sum_{m,n>0} \sum_{\substack{\mathbf{ij}  \\ \mathbf{\mathbf{i}^\prime \mathbf{j}^\prime}}}    { \delta t_{\mathbf{ij}}^\lambda\delta t_{\mathbf{i}^\prime \mathbf{j}^\prime}^\lambda }  \delta\left(E_m+E_n - \hbar \omega_\mathbf{q} \right)\notag \\ & \times (1-f_m-f_n) \beta_{\mathbf{ij} \mathbf{i^\prime j^\prime}}^*\left(\mathbf{q}\right)  \left(  \xi ^{mn}_{{\mathbf{ij}} {\mathbf{\mathbf{i}^\prime \mathbf{j}^\prime}}}(\sigma,\sigma') + \eta ^{mn}_{{\mathbf{\mathbf{i}^\prime \mathbf{j}^\prime}}\mathbf{ij}} (\sigma,\sigma')  \right)  
\label{eq:alpha_4}    \end{align}

where the coherence factors are
\begin{align}
\xi^{mn}_{\mathbf{i}\mathbf{j}\, \mathbf{i}'\mathbf{j}'}(\sigma,\sigma')
&=\sigma \sigma'\left(
u^{m*}_{\mathbf{i}\sigma}\, v^{n*}_{\mathbf{j}\sigma}\,
v^{m}_{\mathbf{i}'\sigma'}\, u^{n}_{\mathbf{j}'\sigma'}
-
u^{m*}_{\mathbf{i}\sigma}\, v^{n*}_{\mathbf{j}\sigma}\,
v^{n}_{\mathbf{i}'\sigma'}\, u^{m}_{\mathbf{j}'\sigma'} \right),
\label{eq:xi_def}
\\[6pt]
\eta^{mn}_{\mathbf{i}\mathbf{j}\, \mathbf{i}'\mathbf{j}'}(\sigma,\sigma')
&=
\sigma \sigma'\left(
v^{m}_{\mathbf{i}\sigma}\, u^{n}_{\mathbf{j}\sigma}\,
u^{m*}_{\mathbf{i}'\sigma'}\, v^{n*}_{\mathbf{j}'\sigma'}
-
v^{m}_{\mathbf{i}\sigma}\, u^{n}_{\mathbf{j}\sigma}\,
u^{n*}_{\mathbf{i}'\sigma'}\, v^{m*}_{\mathbf{j}'\sigma'}\right).
\label{eq:eta_def}
\end{align}
The four contributions, $\alpha_{\mathbf{}}^{\mathrm{I,II,III,IV}}$ (Eqs.~\eqref{eq:alpha_1}--\eqref{eq:alpha_4}) are not independent. They are related by the transformations $(E_m,E_n)\rightarrow(\pm E_m,\pm E_n)$. Exploiting the particle-hole symmetry of the BdG Hamiltonian, these contributions can be mapped onto one another, resulting in a considerably simpler expression for $\alpha_{\mathrm{s}}\left({\mathbf{q},\mathbf{\lambda}}\right)$:

\begin{align}
    \alpha_{\mathrm{s}}\left({\mathbf{q},\mathbf{\lambda}}\right)= &    \frac{\pi  \ell_{\mathbf{q}\mathbf{\lambda}}^2}{{2\hbar v_s N}} \sum_{m,n} \sum_{\substack{\mathbf{ij} \sigma \\ \mathbf{\mathbf{i}^\prime \mathbf{j}^\prime}\sigma'}}    { \delta t_{\mathbf{ij}}^\lambda\delta t_{\mathbf{i}^\prime \mathbf{j}^\prime}^\lambda } \chi_{\mathbf{i j i^\prime j^\prime}}^{mn} (\sigma,\sigma') (f_m-f_n) \notag\\ &  \times   \beta_{\mathbf{ij} \mathbf{i^\prime j^\prime}}\left(\mathbf{q}\right)\delta(E_m-E_n + \hbar \omega_\mathbf{q})
\end{align} 
Note that the sums over (m) and (n) run over the full BdG spectrum rather than being restricted to positive-energy eigenstates. The coherence factors $\chi_{\mathbf{ij}\mathbf{i}^\prime \mathbf{j}^\prime}^{m n} (\sigma , \sigma ')$ with electron-phonon coupling matrix and $\beta$ factors can be further simplified by defining,
\begin{align}
&\mathcal{F}_{mn}^\lambda(\mathbf{q})
=\sum_{\mathbf{ij}\sigma}\delta t_{\mathbf{ij}}^\lambda \phi_{\mathbf{i} \mathbf{j}}^*(\mathbf{q})\left( {u_{\mathbf{i} \sigma}^{m *}} {u_{\mathbf{j} \sigma}^{n }}  -{v_{\mathbf{i} \sigma}^{n }} {v_{\mathbf{j}\sigma}^{m *}} \right) \notag
\end{align}
where $\beta_{\mathbf{ij} \mathbf{i^\prime j^\prime}}\left(\mathbf{q}\right) = \phi_{\mathbf{i} \mathbf{j}}^* \left(\mathbf{q}\right) \phi_{\mathbf{i}^\prime \mathbf{j}^\prime}\left(\mathbf{q}\right)$. Using this, the attenuation coefficient $\alpha_{\mathrm{s}}\left({\mathbf{q},\mathbf{\lambda}}\right)$ can be written as,

\begin{align}
\alpha_{\mathrm{s}}\left({\mathbf{q},\mathbf{\lambda}}\right)= - \frac{\ell_{\mathbf{q}\lambda}^2 }{2 {\hbar v_s N}}\sum_{m,n } |\mathcal{F}_{mn}^\lambda(\mathbf{q}) |^2 \mathrm{Im}[\boldsymbol{\chi}_{\mathbf{q}}^{mn} (\omega_{\mathbf{q}\lambda})]
\end{align}

  where $   \boldsymbol{\chi}_{\mathbf{q}}^{mn} (\omega_{\mathbf{q}\lambda})$ is defined as,
  \begin{align}
       \boldsymbol{\chi}_{\mathbf{q}}^{mn} (\omega_{\mathbf{q}\lambda})
=\frac{
 f(E_m) - f(E_n) 
}{
\hbar \omega_{\mathbf{q}\lambda} + E_m - E_n + i\eta
}
  \end{align}
In our calculation, in addition to the quasiparticle conserving scattering processes we have included the scattering processes $\alpha_{\mathrm{s}}^{\mathrm{IV}}$, where two quasiparticles annihilate together to emit phonons or a phonon is annihilated to create two quasiparticles, though their contribution is negligible for homogeneous superconductors (even for nodal superconductors) across the transition, they become the dominant scattering processes for the bound states scattering, as shown in Fig.~\ref{fig:QPSIII_IV}. The temperature scale ($k_{\mathrm{B}}T$) must remain sufficiently low to prevent thermal smearing of the in-gap bound-state spectrum, whose characteristic energy scale is set by $\omega_q$.
\section{Electron-phonon coupling matrix}
\label{app:2}
The electron-phonon coupling matrix elements $\delta t_{ij}^{\lambda}$
used in the numerical calculations are derived from the first-order
correction to the tight-binding hopping amplitudes introduced in
Sec.~\ref{sec:1}. We assume that the hopping parameter depends only on
the intersite separation $r=|\mathbf r_i-\mathbf r_j|$
[Eq.~\eqref{eqn:ham_1}],
\begin{align}
t\!\left(|\mathbf r_i-\mathbf r_j|\right)\equiv t(r).
\end{align}
Its gradient is directed along the bond connecting sites $i$ and $j$,
\begin{align}
\nabla_{\mathbf r_i} t(r)
=\frac{\partial t}{\partial r}\,\hat{\mathbf r}_{ij},
\qquad
\hat{\mathbf r}_{ij}\equiv
\frac{\mathbf r_i-\mathbf r_j}{|\mathbf r_i-\mathbf r_j|},
\label{eqn:grad_t}
\end{align}
where $\hat{\mathbf r}_{ij}$ is the unit vector along the bond.

To model the distance dependence of the hopping amplitude we assume an
exponential form,
\begin{align}
t(r)=t_0\,e^{-r/\zeta},
\label{eqn:hop_exp}
\end{align}
where $t_0$ sets the hopping scale and $\zeta$ is the decay length.
Evaluated at the equilibrium bond length, the gradient is
\begin{align}
\nabla t(r)\big|_{eq}
=
-\frac{t_{ij}}{\zeta}\,
\hat{\mathbf r}_{0,ij},
\qquad
\hat{\mathbf r}_{0,ij}=
\frac{\mathbf r_{0,i}-\mathbf r_{0,j}}{|\mathbf r_{0,i}-\mathbf r_{0,j}|},
\end{align}
where $t_{ij}$ denotes the hopping amplitudes of nearest-neighbor (nn) and next-nearest-neighbor (nnn) entering the electronic-structure
Hamiltonian. This result is valid for homogeneous regions, but as discussed below it is modified near disorder and inhomogeneities.

The resulting electron-phonon coupling matrix element is then
\begin{align}
\delta t_{ij}^{\lambda}=-
\frac{t_{ij}}{\zeta}\,
\big(\hat{\mathbf{r}}_{0,ij}\cdot\hat{\mathbf e}_{\lambda}\big),
\label{eqn:eph_coupling}
\end{align}
where $\hat{\mathbf{e}}_{\lambda}$ is the polarization vector of phonon
mode $\lambda$. Note that this relation is primarily valid for homogenous systems. For inhomogeneous systems like grain boundaries, the electron-phonon interaction can in principle be much larger than the static electronic hopping parameter. This behavior can arise from local charge redistribution, such as Friedel oscillations near the grain boundary core~\cite{altfeder2012imaging,matsuda2007electron}. Another compelling mechanism leading to a strong enhancement of the electron-phonon interaction near sharp potential barriers involves the explicit dependence of the cross-boundary hopping matrix elements ($t_c$) on the barrier potential \(V(r)\). Following Bardeen’s tunneling theory~\cite{bardeen1961tunnelling}, the effective hopping amplitude connecting the left and right regions across an interface scales inversely with the barrier strength, yielding \(t_c \sim 1/V(x)\). For an atomically sharp, highly localized barrier, the potential profile can be modeled as \(V(x) = V_0 e^{-x/p}\), where \(p \ll 1\) \(\AA\ \)represents a sub-atomic localization length scale. Differentiating these relations with respect to lattice displacement yields \(\nabla t_c / t_c \sim 1/p\), which directly implies the enhancement of coupling near the grain boundary. This demonstrates that the dynamic electron-phonon interaction is strongly amplified by the steep spatial gradient of the localized defect potential. Consequently, in this work, we implement a model featuring an enhanced off-diagonal electron-phonon coupling localized at the disorder boundaries, which successfully bridges and couples electronic states on either side of the grain boundary.

For the tight binding model, we choose $\mathrm{sign}(t)=\mathrm{sign}(t')$, corresponding to the relative sign of the hopping parameters appropriate for the $\mathrm{Sr}_2\mathrm{RuO}_4$ band structure~\cite{walker}. For cuprate superconductors, the nearest-neighbor ($t$) and next-nearest-neighbor ($t'$) hopping amplitudes have opposite signs. However, we find that this difference does not alter the qualitative low-temperature behavior of ultrasonic attenuation.

We have also examined the attenuation for different values of the hopping decay length $\zeta$ and find that the attenuation peak associated with the magnetic bound state remains qualitatively unchanged, provided the electron--phonon coupling across the junction is sufficiently strong and is not overwhelmed by the background contribution.

\begin{figure}[t]
	\centering
	\includegraphics[width=\linewidth]{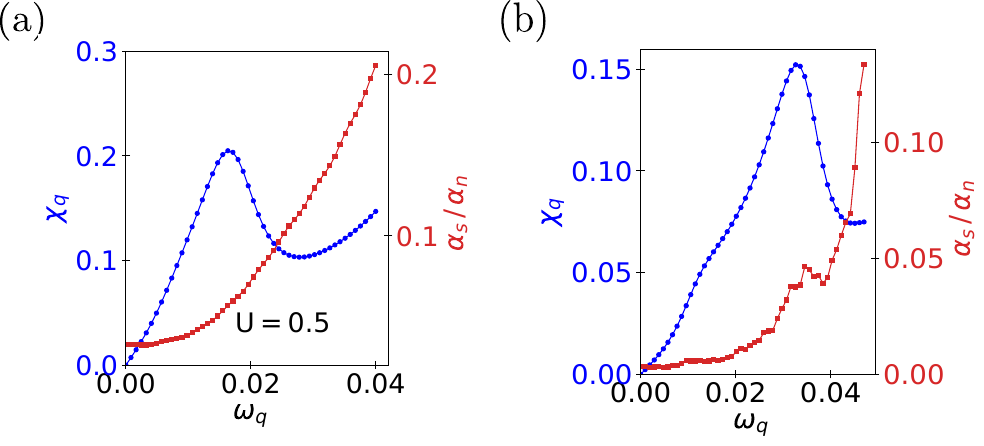
	}
	\caption{Comparison of the ultrasonic attenuation for the magnetic bound state (MBS) case. (a) The tight-binding Hamiltonian is restricted to next-nearest-neighbor hopping, for which no distinct attenuation hump is observed because the short-range electron--phonon interaction does not couple sites across the grain boundary. (b) Weak third-nearest-neighbor hopping and fourth-nearest-neighbor is included, with corresponding  electron--phonon interaction (smaller than used in Fig.~\ref{fig:Chi_alpha_LineImpurity_Vimp}). In this case, the attenuation hump reappears at the bound-state energy, although with reduced intensity because we ignore any local enhancement in the electron--phonon matrix elements and consider an increased transparency of the hopping matrix element across the grain boundary.
	}
	\label{fig:epc_comparison}
\end{figure}

Figure~\ref{fig:epc_comparison} illustrates this behavior by comparing the magnetic bound state case without and with the inclusion of third-nearest-neighbor (3nn) and fourth-nearest-neighbor hopping ($t_{3nn}=-0.075$, $t_{4nn}=-0.025$), which introduces an additional hopping channel across the junction. As shown in Figure~\ref{fig:epc_comparison}(a), the short range electron-phonon interaction does not allow the matrix elements on either side of grain boundary to couple, thus preventing the formation of the hump feature in the attenuation coefficient. However, if we allow both the hopping and the electron-phonon interaction to couple either side of the grain boundary, we find from Figure~\ref{fig:epc_comparison}(b) that the hump feature is restored. We assume in this example that the e-ph matrix element and the cross junction hopping amplitude are both reasonably large, unlike the grain boundary case discussed in Fig.~\ref{fig:Chi_alpha_LineImpurity_Vimp} where only an enhanced e-ph coupling is assumed near the grain boundary. The cross junction hopping makes the interface more transparent, which weakly splits the zero energy bound states, and leads to the spin split bound states (at finite $U$) to move to a higher energy. Another apparent feature of the increase in transparency of the junction is a broadened spectral weight in the susceptibility peak that also contributes to the weakened hump feature at the bound state energies. 
\\
\begin{table}[t]
\centering
\caption{\label{tab:CouplingFactors}
Coupling factors for different sound modes.}
\small
\renewcommand{\arraystretch}{1.3}

\begin{tabular}{|c|c|c|}
\hline
Coupling & L100 & L110 ($b=\sqrt{2}a$)\\
\hline

nn
& $-t$
& $-2t\cos45^\circ\cos(k_yb/2)$ \\
\hline

nnn
& $-2t'\cos45^\circ\cos(k_ya)$
& $-t'$ \\
\hline

3nn
& $-t_{\rm 3nn}$
& $-2t_{\rm 3nn}\cos45^\circ\cos(k_yb)$ \\
\hline
\end{tabular}
\end{table}

%\newpage
% \nocite{*}
% \bibliographystyle{apsrev4-2}
%\bibliography{ref}

%\newpage

\end{document}